\documentclass[letterpaper,11pt]{article}
\pdfoutput=1

\usepackage{jheppub}
\usepackage{multirow}
\usepackage{subfig}
\usepackage{xspace}
\usepackage{xcolor}
\usepackage[countmax]{subfloat}
\usepackage{amsmath}
\usepackage{mathtools}
\usepackage{graphicx}
\usepackage{slashed}
\usepackage{tikz-feynman,contour}
\usepackage{tikz,pgf}
\usepackage{braket}
\usepackage{amsfonts}
\usepackage{amsthm,scrextend}
\usepackage{comment}
\usepackage{placeins}

\newcommand{\as}{\alpha_\text{s}}
\newcommand{\cf}{C_{\text{F}}}
\newcommand{\ca}{C_{\text{A}}}
\newcommand{\tr}{T_{\text{R}}}
\newcommand{\nc}{N_{\text{C}}}

\newcommand{\order}[1]{{\cal O}\left(#1\right)}

\DeclareMathOperator{\De}{d}
\newcommand{\de}{\De\!}

\newcommand{\kt}{k_t}

\newcommand{\mur}{\mu_{\text{R}}}

\usepackage{color}
\definecolor{darkblue}{rgb}{0,0,0.5}
\definecolor{darkgreen}{rgb}{0,0.5,0}
\definecolor{darkorange}{rgb}{0.8,0.3,0}

\newcommand{\pythia}[1]{\textsc{Pythia\xspace #1}}

\title{The Lund $b$-jet plane}

\author[1]{Andrea Ghira,}
\author[2]{Simone Marzani,}
\author[3]{Gregory Soyez}
\affiliation[1]{Technical University of Munich, TUM School of Natural Sciences, Physics Department, James-Franck-Straße 1, 85748 Garching, Germany}
\affiliation[2]{Dipartimento di Fisica, Universit\`a di Genova and INFN, Sezione di Genova,Via Dodecaneso 33, 16146, Italy}
\affiliation[3]{Institut de Physique Th\'{e}orique, Paris Saclay University, CNRS, CEA,
Orme des Merisiers, B\^{a}t 774, F-91191. Gif-sur-Yvette, France}

\emailAdd{andrea.ghira@tum.de}
\emailAdd{simone.marzani@ge.infn.it}
\emailAdd{gregory.soyez@ipht.fr}

\abstract{
We compute the primary Lund plane density for jets initiated by a massive ($b$) quark to single logarithmic accuracy in  Quantum Chromodynamics (QCD).
In order to capture mass effects, we consider quasi-collinear factorisation and we include contributions from the running of the QCD coupling and from collinear evolution, in a variable flavour-number scheme. 
Furthermore, the resummation of soft logarithms, including clustering effects, is performed numerically, keeping the full dependence on the $b$-quark mass. 
While our all-order results can be applied to both hadron and lepton colliders, we present, as first phenomenological application, the resummed calculation of the Lund plane density in $e^+e^-$ collisions at $\sqrt{s}=M_Z$, matched to tree-level matrix elements.
}

\begin{document}
\maketitle

\section{Introduction}\label{sec: intro}

Jets featuring heavy flavours, namely charm ($c$) and beauty ($b$), are of interest for a variety of studies at the Large Hadron Collider (LHC). 
In particular, the development of Infra-Red and Collinear (IRC) safe flavour-jet algorithms~\cite{Banfi:2006hf,Caletti:2022hnc,Czakon:2022wam,Gauld:2022lem,Caola:2023wpj}, recently quantitatively compared in~\cite{Behring:2025ilo}, (for alternative approaches see~\cite{Larkoski:2025afg, Generet:2025gdy}) provides us with the concrete possibility of setting up a successful program of precision physics with flavoured jets.

From an experimental point of view, the lifetime of $B$ (or $D$) hadrons is long enough for their decay to occur away from the interaction point. Dedicated $b$- and $c$-tagging techniques that exploit this property to identify $B$ and $D$ hadrons, or $b$ and $c$ jets, are widely used in collider experiments, see e.g.~\cite{ATLAS:2017bcq, ATLAS:2018nnq,ATLAS:2025dkv}.
From a theory viewpoint, resummed calculations for jets initiated by heavy quarks were first performed in the context of studies focusing on $B$-hadron decays~\cite{Aglietti:2006wh,Aglietti:2007bp,Aglietti:2008xn,Aglietti:2022rcm}
and top jets~\cite{Fleming:2007qr,Fleming:2007xt,Bachu:2020nqn,Jain:2008gb,Hoang:2019fze,Bris:2020uyb}, using the formalism of effective field theory.
Recently, precision studies of heavy flavoured jets have gained the attention of the jet substructure community and an increasing number of phenomenological studies have appeared~\cite{Maltoni:2016ays, Lee:2019lge,Llorente:2014bha,Li:2017wwc,Li:2021gjw, Craft:2022kdo,Cunqueiro:2022svx,Fedkevych:2022mid,Caletti:2023spr,Blok:2023ugf, Zhang:2023jpe,Dhani:2024gtx,Dhani:2025fbk}. 
The LHC experiments have recently measured properties of heavy-flavour jets using a variety of substructure techniques~\cite{ALICE:2025igw,CMS:2024gds,ALICE:2022phr,LHCb:2025tvf}.
One of the most interesting QCD effects affecting the substructure of the heavy-flavour jet is the so-called dead cone~\cite{Dokshitzer:1991fd,Dokshitzer:1995ev}, i.e.\ the suppression of collinear radiation around heavy quarks,   
the first direct observation of which was reported by the ALICE collaboration~\cite{ALICE:2021aqk}.%

In the context of all-orders calculation in QCD, Lund
diagrams~\cite{Andersson:1988gp} provides us with a useful way to
represent the phase space available for the emission of soft and/or collinear gluons off a hard dipole. 
This approach has been proven particularly useful for hadronic final-states resummation, in the presence of multiple scales: it has been successfully applied to the NLL resummation of event shapes~\cite{Banfi:2004yd} of a variety of jet observables, see e.g.~\cite{Marzani:2019hun}. 
Lund diagrams are also extensively exploited in the development of parton-shower algorithms with next-to-leading logarithmic (NLL) accuracy, and beyond, see e.g.~\cite{Dasgupta:2020fwr,vanBeekveld:2024wws}.
Recently, the use of Lund diagrams has been extended to describe the fragmentation properties of heavy quarks~\cite{Ghira:2023bxr,Ghira:2024nkk}.

The Lund jet plane~\cite{Dreyer:2018nbf} is an actual jet substructure observable that stems from Lund diagrams. 
The main conceptual idea behind its construction is that (re-)clustering a
jet with the Cambridge/Aachen (C/A) algorithm~\cite{Wobisch:1998wt, Dokshitzer:1997in} generates a tree of
recombinations that mimics angular ordering and therefore provides a
practical picture similar to the conceptual one underpinning Lund
diagrams.
Therefore, this observable provides us with a theory-inspired representation of a jet and it is therefore amenable to a variety of studies, ranging from classification problems, to studies of QCD properties~\cite{Dreyer:2021hhr,Medves:2022ccw,Medves:2022uii,Baldenegro:2024pfb,ATLAS:2024wrd}. 

The construction of the Lund tree (or Lund jet plane(s)) proceeds as follows. Starting from a regular anti-$k_t$~\cite{Cacciari:2008gp} jet, we recluster its constituents with C/A, which provides us with an angular-ordered tree. 
We then build the Lund tree $\mathcal{L}$ by applying the following
iterative procedure, starting with the full jet $j$:
\begin{enumerate}
\item undo the last step of the clustering to get two subjets
  $j_\text{hard}\equiv j_\text{hard}(j)$ and
  $j_\text{soft}\equiv j_\text{soft}(j)$ from $j$, such that
  $p_{t,\text{hard}}>p_{t,\text{soft}}$;
\item define the following kinematic variables:
  \begin{subequations}\label{eq:lund-kinematic-variables}
  \begin{align}
    \Delta &\equiv \Delta R_{j_\text{hard},j_\text{soft}},& k_t&\equiv p_{t,\text{soft}}\Delta,&
    m^2&\equiv (p_\text{hard}+p_\text{soft})^2,\\
    z&\equiv\frac{p_{t,\text{soft}}}{p_{t,\text{hard}}+p_{t,\text{soft}}}, &\kappa&\equiv z\Delta,&
\psi&\equiv\tan^{-1}\frac{y_\text{soft}-y_\text{hard}}{\phi_\text{soft}-\phi_\text{hard}}.
  \end{align}
  \end{subequations}
  which one can then group into a tuple
  \begin{equation}\label{eq:lund-coordinates}
    \mathcal{T}(j) \equiv\big\{\Delta,k_t,\dots\big\};
  \end{equation}
\item iterate the procedure with $j_\text{hard}$ and $j_\text{soft}$
  (separately). The Lund tree associated with jet $j$ is then defined
  as
  \begin{equation}
    \mathcal{L}(j)\equiv \Big[\mathcal{T}(j), \big(\mathcal{L}(j_\text{hard}), \mathcal{L}(j_\text{soft})\big)\Big].
  \end{equation}
\end{enumerate}
From this full tree of nested kinematic properties associated to each
node of C/A clustering history, we extract the {\it
  primary} Lund plane by keeping only the ones where the iterative
procedure follows the hard subjet. This gives the following set of tuples:
\begin{equation}
  \mathcal{L}_{\text{primary}}(j) \equiv \big[\mathcal{T}(j_1),\dots,\mathcal
  {T}(j_n)\big],
  \quad\text{ with }
  \quad
  \mathcal{T}(j_1)\equiv\mathcal{T}(j)
  \text{ and }
  \mathcal{T}(j_{i+1})\equiv\mathcal{T}(j_\text{hard}(j_i)).
\end{equation}
Similarly, for each jet $j_i$ in the above primary list, one can
define a {\it secondary} Lund plane. Subsidiary planes can be built
similarly at will.
For heavy-quark jets, we will also consider an alternative
declustering procedure where instead of following the hardest branch
at every declustering step, we instead follow the branch to which the
heavy flavour belongs~\cite{Cunqueiro:2018jbh}.

Finally, the construction provided above for a jet produced in hadron
collisions can trivially be extended to $e^+e^-$
collisions.
This implies that the labelling of the hard and soft branches is done based
on the energy of the two subjets and following kinematic variables are
used:\footnote{Note that we have made the choice of using the energy
  of the subjets to define $k_t$ and $z$. Alternatively we could have
  decided to use the 3-momentum instead.}
\begin{align}\label{eq:lund-kinematic-variables-ee}
  \eta &\equiv -\log\tan\frac{\theta_{ij}}{2},&
  k_t&\equiv E_{\text{soft}}\sin\theta_{ij},&
  z&\equiv\frac{E_{\text{soft}}}{E_{\text{hard}}+E_{\text{soft}}}.
\end{align}
In this paper, we consider the primary Lund plane and, in particular, its two-dimensional representation in terms of a set of pairs of kinematic variables defined in Eqs.~
(\ref{eq:lund-kinematic-variables}). We use $(-\log \Delta,\log k_t)$ or, equivalently, $(\eta,\log k_t)$ when considering $e^+e^-$ collisions:
\begin{subequations}
\begin{equation}\label{eq:density-def-delta-kt}
    \rho\left(\Delta, \kt \right)= \frac{1}{N_\text{jets}}\frac{\de^2 n}{\de \log \frac{1}{\Delta} \; \de \log \kt},
\end{equation}
\begin{equation}\label{eq:density-def-eta-kt}
    \widetilde{\rho}\left(\eta, \kt \right)= \frac{1}{N_\text{jets}}\frac{\de^2 n}{\de \eta \; \de \log \kt},
\end{equation}
\end{subequations}
where $n$ is the number of emissions and $N_\text{jets}$ is the number of jets.
Other choices are also common, e.g.\ $(-\log \Delta,-\log z)$.
The primary Lund plane density has been measured at the LHC for several final-state selections~\cite{ATLAS:2020bbn,CMS:2025gdw,CMS:2023lpp,LHCb:2025mcq,Havener:2021yhb}.
In the case of jets initiated by massless partons, the primary Lund
plane density has been computed to single-logarithmic accuracy
in~\cite{Lifson:2020gua}, showing good agreement with the experimental
measurements. The scope of this work is to extend this calculation to the case of jets initiated by massive ($b$) quarks. The structure of the paper closely follows the one of Ref.~\cite{Lifson:2020gua}. For this reason, rather than starting with a summary of the calculation in the case of massless partons, we prefer to discuss in each section similarities and differences of the massive and massless calculations. 

\section{Lund plane density in the quasi-collinear limit}\label{sec:quasi-collinear}

In this section, we present the calculation of the Lund plane density at order $\order{\as}$ exploiting quasi-collinear factorisation~\cite{Catani:2000ef,Catani:2002hc}. Within this approximation, both the transverse momentum \(q_t\) of the emission relative to the emitter and the parton masses \(m_i\) involved in the splitting are small compared to the jet hard scale $Q$,\footnote{The hard scale $Q$ can be thought of as the energy of a jet produced in $e^+e^-$ annihilation or the jet's transverse momentum with respect to the beam axis in $pp$ collisions.} yet they are treated as being of the same order.

We begin by considering, in the quasi-collinear limit, the QCD splitting process \(i \to k(y)+j(1-y)\), and we have 
\begin{align}\label{eq:QC-fact-begin}
    |\mathcal{M}|^2\,  \de \Phi^{(2)} \simeq  |\mathcal{M}_0|^2\frac{\as}{2 \pi}\frac{1}{q_t^2+(1-y)m_k^2+y m_j^2 - y(1-y) m_i^2}P_{ki}(y) \, \de y \, \de q_t^2,
\end{align}
where \(q_t = y(1 - y)\, \theta\, Q\), with \(\theta\) and \(y\) representing the splitting angle and the energy fraction with respect to the emitter, respectively. Moreover, $m_i, m_j, m_k$ are the masses of the particles $i,j,k$ respectively. The unregularised splitting functions $P_{ki}$ are reported in App.~\ref{app:mellin}.
In what follows, 
we will use the generic index $\mathcal{Q}$ for the heavy quark (of
mass $m$) because, although in this paper we will explicitly consider only the case of a $b$ quark, the very same calculation also holds for charm. We instead use $q$ for a massless quark and $g$ for a gluon.

At small angles, the Lund plane density is obtained from Eq.~(\ref{eq:QC-fact-begin}) performing the change of variables:
\begin{align}
      k_t=\frac{q_t}{\max(y,1-y)}, \quad
    \Delta=\frac{k_t}{\min(y,1-y)Q}=\frac{q_t}{y(1-y)Q},
\end{align}
and summing over all the partonic channels. 
The density of the primary plane is constructed by recording the momentum
fraction of the softer branch at each step of the
declustering. Assuming $j$ to be the softest parton, i.e.\ $y>1/2$,
we have
\begin{subequations}\label{eq: rho qc}
\begin{align}
\label{eq: rho qc_a}
    \rho_i^{(\text{q.c.})}(\Delta,\kt)=& \frac{\as}{\pi}\sum_{k=q,\mathcal{Q},g}\frac{y^2(1-y)\, k_t^2}{y^2 k_t^2+(1-y)m_k^2+y m_j^2 - y(1-y) m_i^2}
   P_{ki}(y)\Big |_{1-y= \frac{k_t }{Q \Delta}}\, \\
    = &  \frac{\as}{\pi} z \,\mathcal{P}_i(1-z)\Big |_{z= \frac{k_t }{Q \Delta}},\label{eq: rho qc_b}
\end{align}
\end{subequations}
where we have introduced the Lund plane variable $z=1-y$, and the sum
in Eq.~(\ref{eq: rho qc_a}) runs over all the possible flavours of the
leading parton after the emission (light quark $q$, heavy quark $\mathcal{Q}$ and gluon $g$). Note that, in Eq.~(\ref{eq: rho qc_b}), we have defined, for later convenience,
\begin{align}
\label{eq: P_j}
    \mathcal{P}_i(y)= \sum_{k=q,\mathcal{Q},g}\frac{y^2\, k_t^2}{y^2 k_t^2+(1-y)m_k^2+y m_j^2 - y(1-y) m_i^2} P_{ki}(y).
\end{align}
Explicit expressions for the functions $\mathcal{P}_i,\; i=q, \mathcal{Q},g$, can be easily obtained and they read:
\begin{subequations}\label{eq: splitting P_all}
\begin{align}
\label{eq: splitting P_q}
    \mathcal{P}_q(y)=& P_{qq}(y)+P_{gq}(y), \\
    \mathcal{P}_\mathcal{Q}(y)=& 
    \label{eq: splitting P_Q}
    \frac{y^2 \kt^2}{k_t^2y^2+(1-y)^2 m^2}P_{\mathcal{Q}\mathcal{Q}}(y)+\frac{k_t^2}{m^2+k_t^2} P_{g\mathcal{Q}}(y), \\
     \label{eq: splitting P_g}
    \mathcal{P}_{g}(y)=& 
     P_{gg}(y)+2(n_f-1) P_{qg}(y)+
     \frac{2 y^2\, k_t^2}{y^2 k_t^2+  m^2}
    P_{\mathcal{Q}g}\left(y\right).
\end{align}
\end{subequations}
It is instructive to study the behaviour of the density in the soft
limit. To this purpose, we consider the limit $z\to 0$, i.e.\ $y\to 1$
of Eqs.~(\ref{eq: splitting P_all}), and we find
\begin{subequations}
\begin{align}
\label{eq: rho QC soft}
    \lim_{y\to 1}   \rho_q^{(\text{q.c.})}&= \frac{2\as \cf}{\pi}   + \order{1-y},\\
    \lim_{y\to 1} \rho_\mathcal{Q}^{(\text{q.c.})}&= \frac{2\as \cf}{\pi}\left(\frac{\Delta^2}{\Delta^2+\Delta_d^2} \right)^2 + \order{1-y},\\
    \lim_{y\to 1} \rho_g^{(\text{q.c.})}&= \frac{2\as \ca}{\pi} + \order{1-y} ,
\end{align}
\end{subequations}
where we have introduced
\begin{equation}
  \Delta_d= \frac{m}{Q}
\end{equation}
as the dead-cone angle.

It is interesting to note that, in the soft and collinear limit, the
Lund plane density for light-quark or gluon jets approaches a constant
value, as expected, whereas this behaviour is not observed for jets
initiated by heavy flavours. In the latter case, the Lund plane
density is strongly suppressed by a factor \(\left( \frac{\Delta^2}{\Delta^2 + \Delta_d^2} \right)^2\). This difference is a clear manifestation of the dead-cone effect~\cite{Dokshitzer:1991fd}.

Beyond leading order, the above behaviour receives logarithmic corrections. Each additional factor of \(\as\) is accompanied by up to one logarithm of \(\Delta\), $\Delta_d$, or the ratio \(\frac{Q\Delta}{k_t}\).
Our goal is to calculate this complete set of single-logarithmic contributions to \(\rho(\Delta, k_t)\), outlining the main differences with respect to the massless case.
The logarithmic corrections we want to resum have several physical origins. These are:
\begin{itemize}
    \item[(i)] running coupling corrections, enhanced by logarithms of the transverse momentum \(k_t\);
    \item[(ii)] hard-collinear logarithms of the angles
      $\Delta$ and $\Delta_d$, which induce flavour-changing effects;
    \item[(iii)] soft emissions at large and commensurate angles, enhanced by $\log\left(\frac{Q\Delta}{k_t}\right)$, including the intricate structure of clustering logarithms.
\end{itemize}
We are going to address each of these three contributions in the following sections. 
\section{Running coupling corrections}\label{sec:running-coupling}
Running coupling corrections are universal, i.e.\ they do not depend on the type of collision we are considering, nor on the details of the final state we are selecting. They can be straightforwardly implemented in Eq.~(\ref{eq: rho qc}), following the same procedure adopted in~\cite{Lifson:2020gua} for the massless case:
\begin{equation}
     \rho_i^{(\text{q.c.})}(\Delta,\kt)=  \frac{\as^{\text{CMW}}(k_t^2)}{\pi}  z \, \mathcal{P}_i(1-z)\Big |_{z= \frac{k_t }{Q \Delta}},
\end{equation}
where the suffix CMW indicates that the running coupling is computed with the CMW prescription \cite{Catani:1990rr}:
\begin{align}
  \as^{\text{CMW}}(k_t^2)= \as(k_t^2)\left(1+ \frac{K^{(n_f)}}{2\pi}\as(k_t^2)\right),
\end{align}
with $K^{(n_f)}= \ca \left(\frac{67}{18}-\frac{\pi^2}{6}\right)- \frac{10}{9}n_f\tr$.
Furthermore, we employ the decoupling scheme and, as a prescription to deal with the non-perturbative region, we freeze the coupling below the scale $\Lambda \simeq 1$~GeV:
\begin{align}\label{eq:flv-rc}
    \as(k_t^2)&=\as^{(5)}(k_t^2)\Theta(k_t^2-m^2) +
    \as^{(4)}(k_t^2)\Theta(k_t^2-\Lambda^2) \Theta(m^2-k_t^2) 
   +\as^{(4)}(\Lambda^2)\Theta(\Lambda^2-k_t^2).
\end{align}
In the above equation, $\as^{(n_f)}(k_t^2)$ is the two-loop running
coupling with $n_f$ active flavours:
\begin{subequations}
\begin{align}
    \as^{(5)}(k_t^2)&= \frac{\as}{1+\nu_5}\left(1-\as \frac{\beta_1^{(5)}}{\beta_0^{(5)}}\frac{\log{(1+\nu_5)}}{1+\nu_5}\right),\\
    \as^{(4)}(k_t^2)&=\frac{\as}{1+\nu_4+\delta_{54}}\left(1-\as \frac{\beta_1^{(4)}}{\beta_0^{(4)}} \frac{\log{(1+\nu_4+\delta_{54})}}{1+\nu_4+\delta_{54}}\right)\nonumber \\
    &-\left(\frac{\beta_1^{(5)}}{\beta_0^{(5)}}-\frac{\beta_1^{(4)}}{\beta_0^{(4)}}\right)\log{(1+\nu_{5}^{m})} \frac{\as^2}{(1+\nu_4+\delta_{54})^2},
\end{align}
\end{subequations}
where $\as=\as^{(5)}(Q^2)$ and we have introduced
\begin{align}
        \nu_{n_f}=\as \beta^{(n_f)}_0 \log{\frac{k_t^2}{Q^2}}, \quad 
        \nu^m_{n_f}= \as \beta^{(n_f)}_0 \log{\frac{m^2}{Q^2}}, \quad\text{and}\quad
        \delta_{54}= \nu_{5}^{m}-\nu_{4}^{m}.
\end{align}
This is such that $\as^{(5)}(m^2)=\as^{(5)}(m^2)$.
The two-loop coefficients of the QCD $\beta$-function are
\begin{align}
	\beta_0^{(n_f)}=\frac{11\ca-2 n_f}{12\pi},\qquad \beta_1^{(n_f)}=\frac{17\ca^2-5\ca n_f-3\cf n_f}{24\pi^2}.
\end{align}
Numerical results for running-coupling corrections will be discussed in Sec.~\ref{sec:pheno}.
Next, we turn our attention to the resummation of collinear effects. 

\section{Resummation of collinear logarithms}
\label{sec:collinear}
In the soft and collinear limit, only soft-gluon emissions contribute and, therefore, the flavour of the emitting parton cannot be altered, as explicitly shown in Eq.~(\ref{eq: rho QC soft}). In this limit, all emissions from a quark-initiated jet or a gluon-initiated jet are associated with the colour factor \(\cf\)  or \(\ca\), respectively.
However, at single-logarithmic accuracy, we also have to consider hard-collinear splittings, which can change the flavour of the harder branch. For instance, this happens through a \(q \to qg, \mathcal{Q} \to \mathcal{Q}g\) splitting in which the daughter gluon carries more than half of the momentum of the parent quark, or via a \(g \to q\bar{q}, g \to \mathcal{Q}\bar{\mathcal{Q}}\) splittings.
These effects give rise to collinear logarithms corresponding to a
sequence of emissions strongly ordered in angle and not enhanced by
soft logarithms. This happens, for all flavours, when the angular
resolution scale is above the dead-cone angle, $\Delta > \Delta_d$.
Below this angular scale, the collinear evolution of the heavy flavour
will instead produce logarithms of $\Delta_d$.

Additionally, through collinear evolution, the energy of the leading
parton will decrease, translating into a smearing of the Lund plane
density close to its kinematic boundary. This effect also requires a
resummation of the collinear logarithms.

\subsection{Collinear evolution equations}\label{sec:dglap}
In order to resum collinear logarithms, we introduce, for each flavour
$i$, the function $p_i(x,t)$ as the probability for the leading parton at an angular scale $\Delta$
to have flavour $i$ and energy fraction $x$. This object fulfils a DGLAP-type evolution equation:
\begin{align}
\label{eq: DGLAP type}
    \frac{\de}{\de t}p_i(x,t)= \int^1_0 \frac{\de y}{y} \left(P_{ij}^{(\text{R})}(y) \,p_j\left(\frac{x}{y},t\right)-y P_{ij}
    ^{(\text{V})}(y) \,p_j\left(x,t\right)\right),
    \end{align}
where  the angular dependence is encoded in the  evolution time $t$:
\begin{align}
\label{eq:evolution_time}
t\left(\Delta, \Delta_0\right)= \int^{Q^2 \Delta_0^2}_{Q^2 \Delta^2} \frac{\de k_t^2}{k_t^2} \frac{\as^{\text{CMW}}(k_t^2)}{2\pi},
\end{align}
with $\Delta_0$ being the starting angular scale. Thus, as $t$ increases, we probe smaller angular scales. 
The splitting kernels $P^{(R)}$ and $P^{(V)}$ in Eq.~(\ref{eq: DGLAP
  type}) account for real and virtual emissions, respectively, and
they can be expressed in terms of the standard massless splitting
functions times $\Theta$-functions, which enforce the condition that
at each (real) splitting, we always follow the harder branch. Their explicit expressions are given in App.~\ref{app:mellin}.
We solve the differential equation in Eq.~(\ref{eq: modified dglap}) in two distinct regimes. 
To this purpose, we introduce the dead-cone time
 \begin{align}
    t_d = \int_{Q^2 \Delta_d^2}^{Q^2 \Delta_0^2} \frac{\de k_t^2}{k_t^2} \, \frac{\as^{\text{CMW}}(k_t^2)}{2\pi}.\end{align}
For \( t < t_d \), all parton flavours are evolved with five
massless splitting functions. For \( t > t_d \), the evolution of the
heavy-flavour \( p_\mathcal{Q} \) is frozen. Moreover and the number
of active flavours in the splitting functions is decreased by one.
These conditions reflect the fact that, for splitting angles smaller than \( \Delta_d \), heavy flavours cannot be produced via gluon splittings, nor can they emit gluons at such angles due to the dead cone effect.

We will consider both jets seeded by a heavy flavour $\mathcal{Q}$ and by a light quark $q$. These correspond to different choices of the initial condition of the evolution equation: 
\begin{align}
\label{eq: initial cond}
    p_i(x,0) &= \delta_{ii_0} \delta(1-x),  \quad i_0=\mathcal{Q}, q.
\end{align}
Henceforth, we are going to denote the solution Eq.~\eqref{eq: DGLAP type} with initial condition given by Eq.~\eqref{eq: initial cond} as $p_{i|i_0}(x,t)$.
This solution can be obtained using the numerical method developed in~\cite{Lifson:2020gua} and originally presented in~\cite{Dasgupta:2014yra}. Alternatively, one can solve the equation exactly in Mellin space and subsequently perform a numerical inverse transform to return to \( x \)-space.
This second approach is discussed in detail in App.~\ref{app:mellin}.

For $i_0=\mathcal{Q}$, we also consider the alternative declustering
procedure where one follows the branch to which the heavy flavour belongs.
This avoids contamination from flavour-changing contributions and, furthermore, it greatly simplifies the evolution equation. Indeed, if we consider the heavy-quark initial condition we have $i_0=i=\mathcal{Q}$ and the evolution is the one for the non-singlet distribution, which has no matrix structure:
\begin{align}\label{eq:dglap-non-singlet}
    \frac{\de  }{\de t}p_{\cal Q}(x, t)&=\int_x^1 \frac{\de y}{y}\cf\frac{1+y^2}{1-y}\Big[ p_{\cal Q}\left(\frac{x}{y},t\right)\nonumber -y p_{\cal Q}(x,t) \Big]\nonumber\\
    &=\cf \int_x^1 \frac{\de y}{y} \left(\frac{1+y^2}{1-y} \right)_+ p_{\cal Q}\left(\frac{x}{y},t\right).
\end{align}
At each step of the declustering procedure, we record the kinematic properties of the softer subjet, as done for the standard Lund plane.\footnote{We can instead store the properties of the non-flavoured subjet. In this case we obtain a wider primary plane because the kinematic constraint is $z<1$, rather than $z < \frac{1}{2}$. However, with this choice the comparison to the light-quark case is only qualitative, because we cannot take ratios.} 

\begin{figure}
    \centering
    \includegraphics[width=0.49\linewidth,page=1]{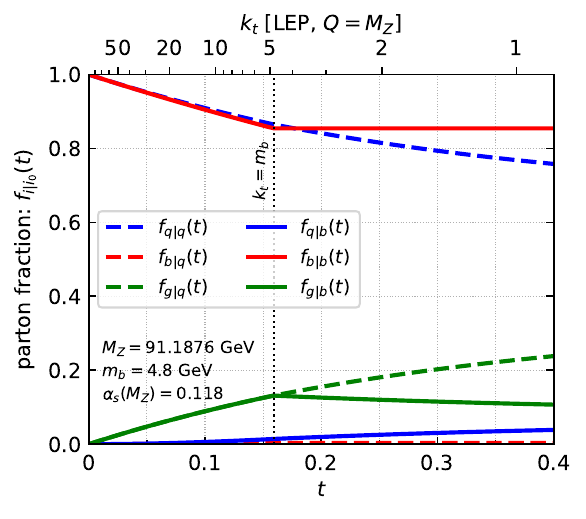} \hfill
    \includegraphics[width=0.49\linewidth,page=3]{figures/partonic-fractions.pdf} 
    \caption{On the left, we show the average partonic fraction $f_{i|i_0}(t)$ assuming the collinear evolution starts from a light quark (dashed lines) or from a heavy quark (solid lines).
    On the right, we show instead the average momentum fraction $\braket{x(t)}_{i|i_0}$ assuming the collinear evolution starts from a light quark (dashed lines) or from a heavy quark (solid lines).
    Collinear evolution is performed by following the hard branch, which corresponds to the standard Lund plane declustering.
    }
    \label{fig:fractions}
\end{figure}

We now study several features of the distributions $p_{i|i_0}(x,t)$. For definiteness, we consider the case of the $b$ as heavy quark, i.e.\ $\mathcal{Q}=b$, with $m_b=4.8$~GeV, and four light flavours. 
We start by considering two integrated quantities, namely the evolution of the average fraction of the different flavours, as well as their longitudinal momentum fraction:
\begin{subequations}
    \begin{align}
    f_{i|i_0}(t) &= \int^1_0\de x \, p_{i|i_0}(x,t),\\ 
    \braket{x(t)}_{i|i_0}&= \frac{1}{f_{i|i_0}(t)}\int^1_0 \de x \, x\, p_{i|i_0}(x,t).
\end{align}
\end{subequations}
These quantities can be computed analytically, because they correspond to the first and second Mellin moments of $p_{i|i_0}(x,t)$, respectively. 
We have computed them in the case of collinear evolution performed by following the hard branch, which corresponds to the standard Lund plane declustering. They are plotted in Fig.~\ref{fig:fractions}.

In Fig.~\ref{fig:fractions}, on the left,  we show the average partonic fraction $f_{i|i_0}(t)$ as function of the collinear evolution time. 
Note that we have added an upper horizontal axis that shows the $\kt$ value probed at any given evolution time, assuming LEP energies. The vertical dotted line indicates the position of the dead cone $t_d$, i.e.\ $\kt=m_b$.
Solid lines correspond to the heavy-quark ($b$) initial condition, while dashed lines correspond to an evolution that start from a light quark. Different colours correspond instead to different partonic flavours at the end of the evolution.
We observe that the fraction of \( b \)-quarks remains constant after the dead-cone time \( t_d \), as expected. This is due to the suppression of gluon radiation for \( t > t_d \), which prevents flavour-changing processes. 

The evolution of the average momentum fraction is instead shown in Fig.~\ref{fig:fractions}, on the right. 
As before, different line styles and colours correspond to different
initial and evolved flavours, respectively. We also show (in black) the average momentum fraction carried by any flavour.  
We observe that for \( t < t_d \), the average momentum fractions carried by light and heavy quarks is the same. This reflects the fact that, in this regime, the jet originates from a quark, and the heavy flavour evolves identically to the light quarks up to \( t_d \).
In contrast, for \( t > t_d \), the heavy quark momentum fraction remains constant, as expected.

\begin{figure}
\centering
\includegraphics[width=0.49\linewidth,page=4]{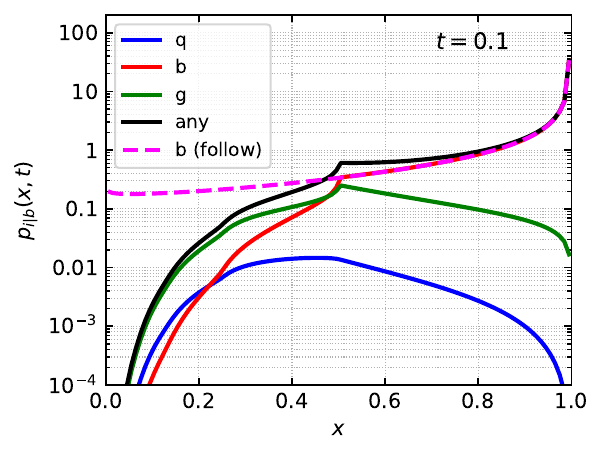}\hfill
\includegraphics[width=0.49\linewidth,page=5]{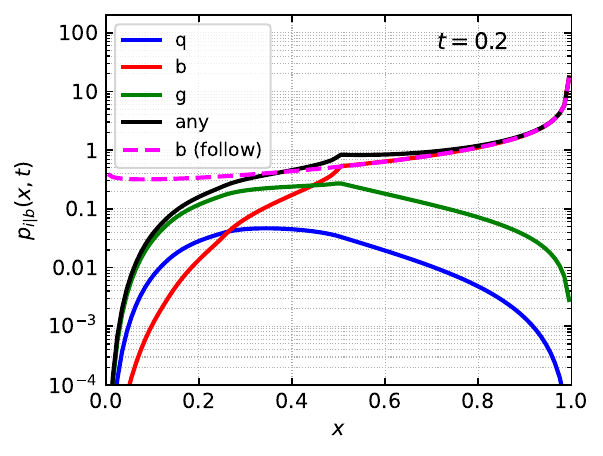}\hfill
\includegraphics[width=0.49\linewidth,page=6]{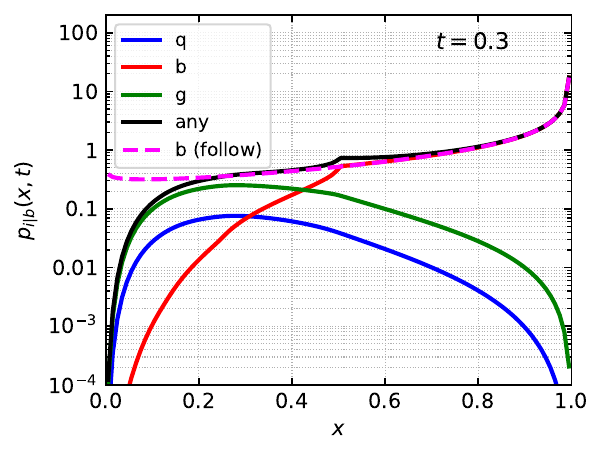}\hfill
\includegraphics[width=0.49\linewidth,page=7]{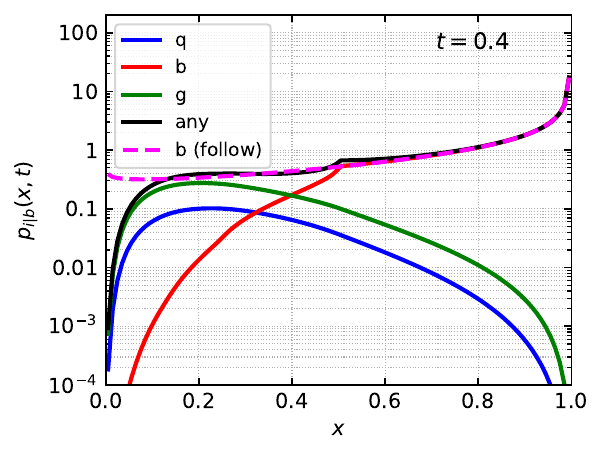}\hfill
    \caption{Plots of the distributions $p_{i|b}(x,t)$ at four different snapshots of the evolution time. In each plot, the different solid lines correspond to a final parton $i$ of a given flavour or any flavour, using standard Lund declustering. The dashed curve instead shows the distribution that corresponds to flavour declustering.}
    \label{fig: dist prob}
\end{figure}

Finally, in Fig.~\ref{fig: dist prob}, we look at the solution of the evolution equations, with the heavy-quark initial condition, i.e.\ $i_0=b$.
The four plots show $p_{i|b}(x,t)$ at four different values of the evolution time ($t=0.1,0.2,0.3,0.4$) as a function of $x$.
In each plot, the different solid lines correspond to a final parton $i$ of a given flavour (blue for $q$, red for $b$, and green for $g$) or any flavour (black) using standard Lund declustering. The dashed curve in magenta instead shows the solution to the non-singlet evolution equation, Eq.~(\ref{eq:dglap-non-singlet}) which implements the follow-the-flavour declustering. 
As expected, at early times, the heavy quark distribution is sharply peaked near $x=1$, but this peak gradually smooths out, as time progresses.
In addition, the gluon and light quark distributions become increasingly enhanced at larger values of $t$ in the small-$x$ region.

\subsection{Collinear Lund plane density}
We are now ready to collect the result we have obtained so far and construct a theoretical prediction for the primary Lund jet plane density that accounts for both running coupling effects and collinear resummation. The collinear density for an initial hard parton \( i_0 \) is given by 
\begin{align}
\label{eq: rho coll}
    \rho_{i_0}^{(\text{coll})}(\Delta, \kt)= \sum_{i=q,\mathcal{Q},g}\frac{\as^{\text{CMW}}(k_t^2)}{\pi}\int^1_0 \de x\, p_{i|i_0}(x, t(\Delta,\Delta_0)) \, z \, \mathcal{P}_i(1-z)\Theta(1-2z)\Big|_{z= \frac{k_t }{x Q \Delta}}.
\end{align}
In the above expression, the distribution $p_{i|i_0}$, which accounts for the collinear evolution from parton $i_0$ at the initial angular scale $\Delta_0$, to parton $i$ at angular resolution $\Delta$, is convoluted with the quasi-collinear splittings $\mathcal{P}_i$, previously defined in Eq.~(\ref{eq: P_j}).
Numerical results for the collinear contribution to the density are presented in Sec.~\ref{sec:pheno}.

\section{Resummation of soft logarithms}
\label{sec:soft}
We now turn our attention to the resummation of soft logarithms of the form $\log \frac{Q \Delta}{k_t}=\log\frac{1}{z}$. These contributions only arise from emissions of soft gluons, which cannot alter the flavour of the emitters. However, soft emissions at large angles are sensitive to the colour structure of the hard scattering, giving rise to a rather complex resummation structure. Following what was done for the massless calculation~\cite{Lifson:2020gua}, we perform this resummation numerically by means of a dipole shower, in the large $\nc$ limit. 
We will describe this numerical approach in Sec.~\ref{sec:num_soft_res}. Prior to that, it is useful to consider soft emissions at first and second order, in the strong coupling. 

\subsection{Soft emissions at small angles}
In this section we perform a fixed-order analysis of soft gluon emissions. 
We work in the small-$\Delta$ limit. This allows us to perform most of the integrals analytically and it is enough to shed light on the intricate structure of clustering logarithms.
We will reinstate the full angular dependence when considering the full (numerical) resummation of soft effects in Sect.~\ref{sec:num_soft_res}.

\subsubsection{$\order{\as}$ calculation}\label{sec:soft-small-angle-as}

    Let us start by considering the emission of a soft gluon with momentum $k$ off a  dipole with momenta $p_a$ and $p_b$.
In the soft limit, the squared matrix element takes the well known expression: 
\begin{equation}\label{eq:w-massive}
    w^{ab,l}= \frac{p_a\cdot p_b}{p_a\cdot k_l \, p_b\cdot k_l}-\frac{m_a^2}{2 (p_a \cdot k_l)^2}-\frac{m_b^2}{2 (p_b \cdot k_l)^2}.
\end{equation}
with $m_a, m_b$ the masses of the emitting particles. Details of the dipoles' calculations are given in App.~\ref{app:dipoles}.
Here, we focus our attention on a dipole which has at least a hard parton in final-state and it therefore initiates the jet we are considering. As previously mentioned, we work in the small angle limit.
In such an approximation, we can assign a flavour label to the dipole, i.e. $w^{ab,l}\to w^{ab,l}_i, i=q, \mathcal{Q}, g$.
The soft (and small-angle) contribution to the Lund plane density at $\order{\as}$ is
\begin{align}
\label{eq:rho-soft-one-emission-simple}
    \rho^{\text{(soft)}}_{i}= \frac{\as}{2\pi} C_i
    \int_0^{Q} \omega \de \omega \int \de \theta^2 \int_0^{2 \pi}\frac{\de \phi}{2 \pi}  w_i^{ab,l} \Delta \delta(\theta-\Delta) k_t \delta( k_t - \omega \Delta),
\end{align}
where $C_i$ is the appropriate colour factor.
Calculating the integral in Eq.~(\ref{eq:rho-soft-one-emission-simple}), we obtain:
    \begin{align}
     \rho^{\text{(soft)}}_{q}= \frac{2 \as \cf}{\pi}, \quad  
      \rho^{\text{(soft)}}_{\cal Q}= \frac{2 \as \cf} {\pi} \left(\frac{\Delta^2}{\Delta^2+\Delta_d^2}\right)^2, \quad
       \rho^{\text{(soft)}}_{g}= \frac{2 \as \ca}{\pi}.
\end{align}
As expected, in the collinear limit, the calculation at order $\as$ coincides with the soft limit of the quasi collinear approximation shown in Eq.~\eqref{eq: rho QC soft}.
\subsubsection{$\order{\as^2}$ calculation}\label{sec:soft-small-angle-as2}
We now move to the $\mathcal{O}(\as^2)$ contribution. 
We work in the limit in which the energies are strongly-ordered, $\omega_2\ll \omega_1\ll Q$, and the angles are small but commensurate, $\theta_1\sim\theta_2\sim\theta_{12}\ll 1$. These approximations allow us to study  structure of clustering logarithms.

We begin by considering a jet initiated by a massless parton, either a quark or a gluon. These results were previously derived in~\cite{Lifson:2020gua}. We briefly review them here, in order to highlight similarities and differences with the heavy-quark case. 
We analyse the squared matrix element for the emission of two soft gluons, strongly ordered in their energy from two massless quark legs. The squared matrix element for this process was originally computed in~\cite{Catani:1983bz,Dokshitzer:1992ip}, and is given by:
\begin{align} \label{eq:double-soft-massless}
    W_q &= 2 \cf w_q^{ab,1}\left[\ca w_q^{a1,2}+\ca w_q^{b1,2}+\left(2 \cf -\ca\right)w_q^{ab,2}\right]\\
    &= 4\cf^2 \,   w_q^{ab,1} w_q^{ab,2} + 2 \cf \ca\, w_q^{ab,1}
    \left(w_q^{a1,2}+ w_q^{b1,2}-w_q^{ab,2}\right)= \cf^2 W_{q,\cf^2}+ \cf \ca W_{q,\cf \ca},\nonumber
\end{align}
where the $\cf^2$ term describes the independent emission contribution (Abelian contribution), while the $\cf \ca$ one, the correlated (non-Abelian) one. As before, we have assigned flavour labels to the dipoles because we are going to consider the small angle limit. 

Let us first consider the independent-emission case and compute the
contribution to the primary Lund plane density arising from gluon
$2$.\footnote{In the strongly-ordered limit, the contribution from the
  first gluon cancels between the real and virtual terms.} In the small angle limit, we have:
\begin{align}\label{eq:clust-initial-expression-cf2}
\rho_{q,\cf^2}^{(\text{soft})}
=& \left(\frac{\as \cf}{4\pi}\right)^2
  \int_0^{Q}\omega_1 \de \omega_1
  \int_0^{\omega_1}\omega_2\de \omega_2
  \int^1_0 \de \theta^2_1   \int^1_{0} \de \theta^2_2 \int^{2\pi}_0 \frac{\de \phi}{2\pi} W_{q,\cf^2}
  \nonumber\\&
  \Delta \, \delta(\Delta-\theta_2) \, k_t \, \delta(k_t-\omega_2\Delta) 
    \left\{\left[
      1-\Theta(\theta_{12}<\theta_1)\Theta(\theta_{12}<\theta_2) \right]-1\right\}.
\end{align}
The first two contributions correspond to the double-real emission, with the requirement that gluons $1,2$ are not recombined together (otherwise they would belong to the secondary plane) and the third one is the real-virtual correction. Using $\theta^2_{12}= \theta_1^2+\theta_2^2-2 \theta_1 \theta_2 \cos \phi$ and performing the integrals over $\theta_2$ and $\omega_2$ with the $\delta$ functions, we arrive at
\begin{align}\label{eq:clust-cntd-expression-cf2}
\rho_{q,\cf^2}^{(\text{soft})}&= \left(\frac{2\as \cf}{\pi}\right)^2   \frac{\text{Li}_2\left (e^\frac{i \pi}{3}\right) -\text{Li}_2\left (-e^\frac{ i 2 \pi}{3}\right)}{2 \pi i}\log\frac{k_t}{Q \Delta}+\order{\as^3}\nonumber\\
\simeq&
  \left(\frac{2\alpha_s \cf}{\pi}\right)^2  0.323066 \log\frac{k_t}{Q \Delta}+\order{\as^3}.
\end{align}
This result is in full (numerical) agreement with the one found in~\cite{Lifson:2020gua}, however we find a simpler analytic expression for the coefficient of the logarithm.

Next, we look at the correlated emissions. Here, the
contribution from gluon 2 only comes from configuration with two real
emissions. Thus, we have:
\begin{align}\label{eq:clust-initial-expression-cfca}
\rho_{q,\cf \ca}^{(\text{soft})}
=& \left(\frac{\alpha_s}{4\pi}\right)^2\cf \ca
  \int_0^{Q}\omega_1 \de \omega _1 \int^{\omega_1}_{0} \omega_2 \de \omega_2 \int^{1}_{0} \de \theta^2_1\int^{1}_0 \de \theta^2_2 \int^{2\pi}_0 \frac{\de \phi}{2\pi} W_{q,\cf \ca}\nonumber\\&
  \Delta \, \delta(\Delta-\theta_2) \, k_t \, \delta(k_t-\omega_2\Delta) 
    \left[
      1-\Theta(\theta_{12}<\theta_1)\Theta(\theta_{12}<\theta_2)
  \right].
\end{align}
 Following the same steps as before, we arrive at
\begin{align}\label{eq:clust-cntd-3-expression-cfca}
\rho^{(\text{soft})}_{q,\cf \ca}
=
-\left(\frac{2\alpha_s}{\pi}\right)^2\cf \ca
\log \frac{k_t}{Q \Delta} \left[
  \int_0^\frac{\pi}{6} \frac{\de \phi}{\pi} \log \left(4\cos^2 \phi\right)- \int_\frac{\pi}{6}^\frac{\pi}{2} \frac{\de \phi}{\pi} 
  \log\left(2\cos\phi\right)\right].
\end{align}
The final angular integral yields (rather surprisingly) the same numerical coefficient as in the independent emission case, so the soft contribution to the Lund plane density in the collinear limit becomes:
\begin{align}\label{eq:final2gluons-massless}
\rho_q^{(\text{soft})}  =&\frac{2 \as \cf}{\pi}+0.323066
  \left(\frac{2\alpha_s} {\pi}\right)^2 \cf(\cf- \ca) \log\frac{k_t}{Q \Delta}+\order {\as^3} \nonumber \\
  =& \frac{2 \as \cf}{\pi}\bar{\rho}_q^{(\text{soft})}.
\end{align}
We can repeat the same analysis for gluon jets. This calculation simply reduces to perform the substitution $\cf\to \ca$. Thus it follows that
\begin{align}\label{eq:final2gluons-gluon}
\rho_g^{(\text{soft})}  =\frac{2 \as \ca}{\pi}+\order {\as^3} = \frac{2 \as \ca}{\pi}\bar{\rho}_g^{(\text{soft})}.
\end{align}
The factors $\bar{\rho}_q^{(\text{soft})}$, $ \bar{\rho}_g^{(\text{soft})}$ were implicitly defined for later convenience.

We then consider the case of the emission of two soft gluons off a
massive quark-antiquark dipole. The squared amplitude for this process
was computed in Refs.~\cite{Czakon:2011ve,Czakon:2014oma}. Notably, in
the strongly ordered limit, the massive squared matrix element retains
the same structure as Eq.~(\ref{eq:double-soft-massless}), with the
massless eikonal factor $w_q^{ab,l}$ replaced by the massive version
$w_{\mathcal{Q}}^{ab,l}$. We choose $p_a$, see
Eq.~(\ref{eq:double-soft-massless}), to be the momentum of the heavy
quark that initiates the jet, and we work in the quasi-collinear limit
with respect to its momentum direction. Therefore, we can set
$m_b=0$. In the quasi-collinear limit, the Abelian and non-Abelian
contributions to the squared matrix element read
\begin{subequations}
\begin{align}
    W_{\mathcal{Q},\cf^2}=& \frac{16}{\omega_1^2 \omega_2^2} \frac{\theta_1^2 \theta_2^2}{(\theta_1^2+\Delta_d^2)^2(\theta_2^2+\Delta_d^2)^2}, \\
    W_{\mathcal{Q},\cf \ca}=& \frac{16}{\omega_1^2 \omega_2^2} \frac{\theta_1^2}{(\theta_1^2+\Delta_d^2)^2}\frac{\Delta_d^2+\theta_1\theta_2 \cos \phi}{(\theta_2^2+\Delta_d^2)(\theta_1^2+\theta_2^2-2 \theta_1 \theta_2 \cos \phi)}.
\end{align}
\end{subequations}
We integrate these contributions with the same constraints as in massless case.
We start by considering the massive calculation for the $\cf^2$
contribution. As before, we take the quasi-collinear emission of the
soft matrix element
\begin{align}\label{eq:clust-initial-expression-cf2-massive}
\rho^{(\text{soft})}_{\mathcal{Q},\cf^2}
=& \left(\frac{\as \cf}{4\pi}\right)^2
  \int_0^{Q}\omega_1 \de \omega_1
  \int_0^{\omega_1} \omega_2 \de \omega_2
  \int \de \theta^2_1
  \int \de \theta^2_2 \int^{2\pi}_0 \frac{\de \phi}{2\pi}\;W_{\mathcal{Q},\cf^2}
  \nonumber\\&
  \Delta \, \delta(\Delta-\theta_2) \, k_t \, \delta(k_t-\omega_2\Delta) 
    \left\{\left[
      1-\Theta(\theta_{12}<\theta_1)\Theta(\theta_{12}<\theta_2)\right]
  -1\right\}
\end{align}
Proceeding as in the massless case, we find
\begin{align}\label{eq:clust-cntd-expression-cf2-massive}
\rho^{(\text{soft})}_{\mathcal{Q},\cf^2}&=\left(\frac{2\alpha_s \cf}{\pi}\right)^2 \log\frac{k_t}{Q \Delta}\left(\frac{\Delta^2}{\Delta^2+ \Delta_d^2}\right)^2 \mathcal{F}_{\cf^2}\left(\frac{\Delta_d^2}{\Delta^2}\right),
\end{align}
where:
\begin{align}\label{eq: FA}
    \mathcal{F}_{\cf^2}\left(x\right)= \int_0^\frac{\pi}{3}\frac{\de \phi}{2 \pi} \Bigg[\log \frac{4 \cos^2 \phi\; (4 \cos^2 \phi +x)}{1+ 4 x \cos^2 \phi }
 +\frac{x(1-16 \cos^4 \phi)}{(4 \cos^2 \phi +x)(1+ 4 x \cos^2 \phi)}
\Bigg].
\end{align}
Equation (\ref{eq:clust-cntd-expression-cf2-massive}) reduces to the massless result, Eq.~(\ref{eq:clust-cntd-expression-cf2}) when $\Delta_d\to 0$. Note that in the opposite limit, $ \Delta_d \gg \Delta$, both the prefactor $\left(\frac{\Delta^2}{\Delta^2+\Delta_d^2}\right)^2$ and $\mathcal{F}_{\cf^2}$ vanish, giving an overall $\left(\frac{\Delta^2}{\Delta^2+\Delta_d^2}\right)^4$ suppression. This is expected because, in this limit, the two emissions are independent and they both feel the dead cone. This results into a suppression factor which is the square of the one-emission contribution.

We now move to the correlated emission contribution:
\begin{align}\label{eq:clust-initial-expression-cfca-massive}
\rho^{(\text{soft})}_{\mathcal{Q},\cf\ca}
&= \left(\frac{\as}{4\pi}\right)^2\cf \ca
  \int_0^{Q} \omega_1 \de \omega_1
  \int_0^{\omega_2}\omega_2\de \omega_2
  \int \de \theta_1^2
  \int \de \theta_2^2 \int^{2\pi}_{0} \frac{\de \phi}{2\pi} W_{\mathcal{Q},\cf \ca}
  \nonumber\\&
  \Delta \, \delta(\Delta-\theta_2) \, k_t \, \delta(k_t-k_{\perp 2}\Delta) 
    \left[
      1-\Theta(\theta_{12}<\theta_1)\Theta(\theta_{12}<\theta_2)
    \right]
\end{align}
Performing the azimuthal integral first, we arrive at
\begin{align}
\label{eq: rho cf ca }
    \rho_{\mathcal{Q}, \cf \ca}^{(\text{soft})}= -\left(\frac{2 \as}{\pi}\right)^2\cf \ca \log \frac{\kt}{Q \Delta}\left(\frac{\Delta^2}{\Delta^2+\Delta_d^2}\right)^2\mathcal{F}_{\cf\ca}\left(\frac{\Delta^2_d}{\Delta^2}\right),
\end{align}
where the correlated-emission form factor is defined as:
\begin{align}
\label{eq: FNA}
    \mathcal{F}_{\cf \ca}(x)=& 
   \int_0^\infty \de y \,  \frac{(1+x)y^3}{(y^2+x)^2} \Bigg[ \frac{\min(1, y^2)+x}{|1-y^2|} \nonumber  \\
   +&
   \frac{\phi_0(1-y^2)-2 (1+2 x +y^2)\tan^{-1}\left(\frac{1+y}{1-y}\tan \frac{\phi_0}{2}\right)}{2 \pi (1-y^2)}
\Theta\left(2y-1\right)\Theta(2-y)\Bigg].
\end{align}
The integral in this expression is over $y= \frac{\theta_1}{\Delta}$ and we have introduced $\phi_0= \cos^{-1}\max  \left(\frac{ y}{2} , \frac{1}{2 y}\right)$.
It is easy to check that non-Abelian form factor correctly reduces to the massless result in the limit $\Delta_d \to 0$. Studying the opposite limit, $\Delta_d \gg \Delta$, is less straightforward.

\begin{figure}
    \centering
    \includegraphics[width=0.49\linewidth]{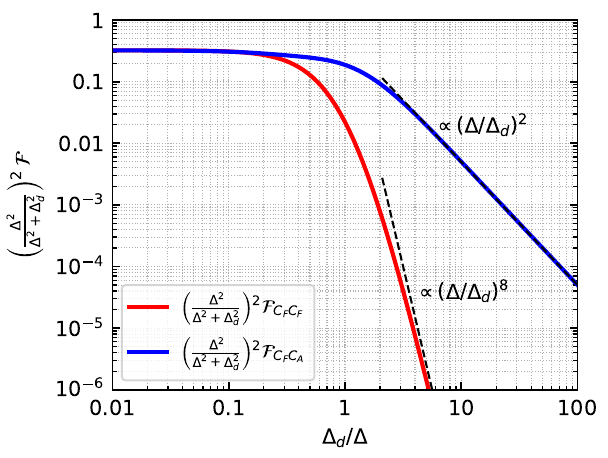}\hfill
    \includegraphics[width=0.49\linewidth,page=1]{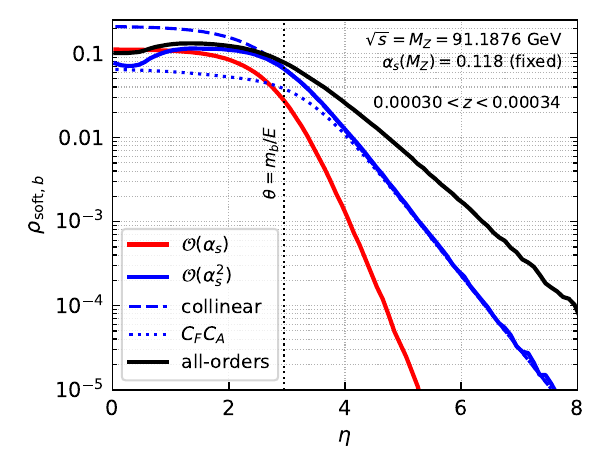}
    \caption{On the left, comparison between the Abelian and non-Abelian contributions to $\rho_{\mathcal{Q}}^{(\text{soft})}$. The Abelian contribution leads to a stronger suppression than the non-Abelian one, as discussed in the text. On the right, all-order resummation of soft contributions for a $b$-quark dipole ($\mathcal{Q}=b$) obtained with the numerical method described in Sec.~\ref{sec:num_soft_res}, as well as its first and second order expansions, plotted as a function of the Lund-plane pseudorapidity $\eta=-\log \tan \frac{\theta_{ij}}{2}$, see Eq.~(\ref{eq:lund-kinematic-variables-ee}).
    The second order expansion is compared to the analytic calculation performed in the collinear limit.}
    \label{fig: rho abelian and non abelian}
\end{figure}

We begin with a numerical study.
We show in Fig.~\ref{fig: rho abelian and non abelian}, the Abelian and non-Abelian contributions to the density, Eqs.~(\ref{eq:clust-cntd-expression-cf2-massive}) and ~(\ref{eq: rho cf ca }), each normalised to the respective massless case, i.e.\ $\rho^{(\text{soft})}_{\mathcal{Q},\cf^2}/\rho^{(\text{soft})}_{q,\cf^2}$  and $\rho^{(\text{soft})}_{\mathcal{Q},\cf \ca}/\rho^{(\text{soft})}_{q,\cf \ca}$.
These ratios only depend on $\Delta$ through the scaling variable
$x=\left(\Delta_d/\Delta\right)^2$ and approach unity when $x\to
0$. The plot shows striking differences in behaviour between the two contributions. The Abelian one (in red) exhibits a rather strong suppression in the deep dead-cone region. We have already commented on this: the two soft gluons are emitted independently and, consequently, the suppression factors multiply. 
The non-Abelian contribution shows instead less suppression. Qualitatively, this is not surprising. Indeed, in the correlated-emission the two soft gluons are not independent and the soft logarithm arises from the emission of the softest gluon off a dipole formed by the heavy quark and the other gluon. Thus, in this case we would therefore expect a suppression factor equal to the one-gluon emission case, i.e.\ $\left(\frac{\Delta^2}{\Delta^2+\Delta_d^2}\right)^2=\left(1+x\right)^{-2}\sim x^{-2}$, at large $x$.
However, a closer inspection of Fig.~\ref{fig: rho abelian and non abelian} reveals that the actual scaling of the non-Abelian is actually $x^{-1}$.
This behaviour is somewhat counterintuitive, as it implies that the correlated emission of two soft gluons is less suppressed than the emission of a single gluon.

In order to understand the origin of this behaviour, let us go back to Eq.~(\ref{eq: FNA}). 
We first note that at any finite $y$, the integrand is constant at large $x$. Thus, any alteration to the one-gluon suppression term that appears in Eq.~(\ref{eq: rho cf ca }) must originate from the large-$y$ behaviour. 
Consequently, we note that the leading contribution cannot come from the second line of Eq.~(\ref{eq: FNA}) because the $y$-integration is bounded. 
In contrast, the behaviour of the integral in the first line of Eq.~(\ref{eq: FNA}) has the following asymptotic behaviour
\begin{align}\label{eq:asympotic_F_NA}
    \mathcal{F}_{\cf \ca}(x) =\frac{x}{2} +\order{x^{-1}}.
\end{align}
When we substitute this result into Eq.~(\ref{eq: rho cf ca }), we find the following asympotic behaviour for the correlated-emission contribution to the density
\begin{align}
\label{eq: rho cf ca asymptotic }
    \rho_{\mathcal{Q}, \cf \ca}^{(\text{soft})}= 
    \frac{1}{2}\left(\frac{2 \as}{\pi}\right)^2\cf \ca \left(\frac{\Delta^2}{\Delta^2+\Delta_d^2}\right) \log \frac{\kt}{Q \Delta}+ \dots,
\end{align}
which is indeed less suppressed than the single-emission contribution. 

It is worth noting that the behaviour found in Eq.~(\ref{eq:asympotic_F_NA}) arises from the integration region where $y^2 \sim x \gg 1$, which
corresponds to $\theta_1 \sim \Delta_d \gg \theta_2 = \Delta$. That is, one gluon is emitted at an angle that is commensurate with the dead-cone angle, while the other gluon goes deep into the dead-cone region.
To gain further insight into the properties of this contribution, we express the emission probability of correlated gluon emission in terms of angular variables
\begin{align}
\label{eq: double differential}
    \frac{\de^2 \mathcal{P}}{\de \log \theta^2_1 \de \log \theta^2_2}=  \frac{\theta_2^2}{|\theta_1^2-\theta_2^2|} \left(\frac{\theta_1^2}{\theta_1^2+\Delta_d^2}\right)^2 \frac{\min(\theta_1^2, \theta_2^2)+\Delta_d^2}{\theta_2^2+\Delta_d^2},
\end{align}
which modulo prefactors corresponds to the integrand in the first line of Eq.~\eqref{eq: FNA}, if we identify $y^2= \frac{\theta_1^2}{\theta_2^2}$, and $x= \frac{\Delta_d^2}{\theta_2^2}$.
We note that Eq.~(\ref{eq: double differential}) is written as the product of the single emission contribution  $\left(\frac{\theta_1^2}{\theta_1^2+\Delta_d^2}\right)^2$ times the massless factor $\frac{\theta_2^2}{|\theta_1^2-\theta_2^2|}$ which originates non-global logarithms~\cite{Dasgupta:2001sh}.
This is a non-trivial observation: the non-Abelian contribution from a
first gluon emission at $\theta_1\sim \Delta_d$ extends arbitrarily
deep inside the primary plane ($\Delta\ll\Delta_d$) and this
contribution is less suppressed than the Abelian one.\footnote{It
  appears that the dead-cone introduces an angular cut-off that gives
  rise to a non-global configuration, similarly to what happens with
  sharp phase-space boundaries or jet definitions. (SM would like to
  thank Yacine Mehtar-Tani for interesting discussions about this
  topic.)}
Anticipating our all-order discussion below, we think that this
behaviour extends beyond two emissions and that the dominant
contribution in the resummed density for $\Delta\ll\Delta_d$ would
come from a coherent emission from an arbitrary number of more
energetic (correlated) gluons at angles commensurate to $\Delta_d$.

Finally, as done for the case of a jet initiated by a massless quark, Eq.~(\ref{eq:final2gluons-massless}), or by a gluon, Eq.~(\ref{eq:final2gluons-gluon}), we report the expression for the soft- and small-angle contribution to the primary Lund plane density of a heavy quark
\begin{align}\label{eq:final2gluons-HQ}
    \rho^{(\text{soft})}_{\mathcal{Q}}&=
    \frac{2\as \cf}{\pi} \left(\frac{\Delta^2}{\Delta^2+\Delta_d^2}\right)^2+ \left(\frac{2\as}{\pi}\right)^2\cf\left[\cf\mathcal{F}_{\cf2}\left(\frac{\Delta_d^2}{\Delta^2}\right)-\ca\mathcal{F}_{\cf \ca}\left(\frac{\Delta_d^2}{\Delta^2}\right)\right]\nonumber \\
   &\quad \times\left(\frac{\Delta^2}{\Delta^2+\Delta_d^2}\right)^2 \log \frac{\kt}{Q\Delta} +\order{\as^3}.
   \nonumber \\&= \frac{2 \as \cf}{\pi}\left(\frac{\Delta^2}{\Delta^2+\Delta_d^2}\right)^2 \bar \rho^{(\text{soft})}_{\mathcal{Q}},
\end{align}
where, as before, $\bar \rho^{(\text{soft})}_{\mathcal{Q}}$ has been defined for future convenience. 
In the next section we discuss how to generalise the above result to all orders and beyond the small-angle approximation.

\subsection{Soft emissions to all orders}\label{sec:num_soft_res}

The resummation of soft emission at wide or commensurate angles is notoriously a difficult problem, even at single-logarithmic accuracy and analytic approaches exist only in limited cases, see e.g.~\cite{Banfi:2002hw}.
The method followed in~\cite{Lifson:2020gua} takes inspiration from the resummation of non-global ~\cite{Dasgupta:2001sh} and clustering logarithms~\cite{Appleby:2003sj,Banfi:2005gj}.
Namely, we work in the large-$\nc$ limit and we construct the all-order result through a dipole shower ordered in the transverse momentum of the emissions. 

To achieve this, we start with a Born event and we decompose it, in the large-$\nc$ limit into a series of dipoles ($i,j$), each with appropriate weight $w_{ij}$. For a given initial dipole ($i,j$), we generate the emission of a soft gluon $k$ with probability given by the eikonal factor. The procedure is then repeated for each of the daughter dipoles $(i,k)$ and $(j,k)$, and so on, until one reaches a cutoff scale. 
We preform this operation for each initial dipole in the event. The
corresponding event (including all the particles in the event) is then
clustered to produced the Lund plane density.
In general, a given process can receive contributions from different
initial dipole configurations, i.e.\ colour flows $\mathcal{C}_i$, each with a weight
$w_i$, that need to be summed over:
\begin{align}\label{eq:rho_soft_dipole_sum}
    \rho^{(\text{soft})}(\Delta,\kt)&= \sum_{\substack{\text{partonic}
  \\ \text{channel}}} \sum_{\mathcal{C}_i} w_{i}\,
    \left[\rho^{(\text{soft})}(\Delta,\kt)\right]_{\mathcal{C}_i}.
\end{align}
Note that because the dipole evolution only allows for the emission of soft gluons, flavour can never change in the soft sector, as opposed to the collinear one. 
Thus, the only new ingredient that we have to account for is the
possibility that the initial hard dipoles can feature one or two massive legs. This affects the resummation in two ways. First, the eikonal factor that controls the emission probability becomes mass-dependent. Second, reconstruction of the kinematics of soft gluons emitted from dipoles with massive leg(s) is slightly altered. Both effects are easy to account for and details are given in App.~\ref{app:dipoles}.

From a technical point of view, the evolution of the dipoles is performed at fixed-coupling.
That is, from the dipole shower we obtain the Lund plane coordinates $(\Delta,z)$ and hence the fixed-coupling density in the soft limit as $\rho^{(\text{soft})}(\Delta,z Q \Delta)$.
To account for running-coupling eﬀects, we follow~\cite{Lifson:2020gua} and determine an effective energy fraction $z_\text{eff}$ as
\begin{equation}
   z_\text{eff}= \exp \left[ -\frac{\pi}{\as}t\left(\kt,  Q \Delta\right)\right],
\end{equation}
where the evolution time is given by the integral over the running coupling, see Eq.~(\ref{eq:evolution_time}). We then evaluate the soft-resummed density with the effective energy fraction, $z_\text{eff}$, i.e.\
\begin{equation}
\rho^{(\text{soft})}(\Delta,z Q \Delta)
\to \rho^{(\text{soft})}(\Delta,z_\text{eff} \, Q \Delta).
\end{equation}

\begin{figure}
\centering
\includegraphics[width=0.49\linewidth,page=5]{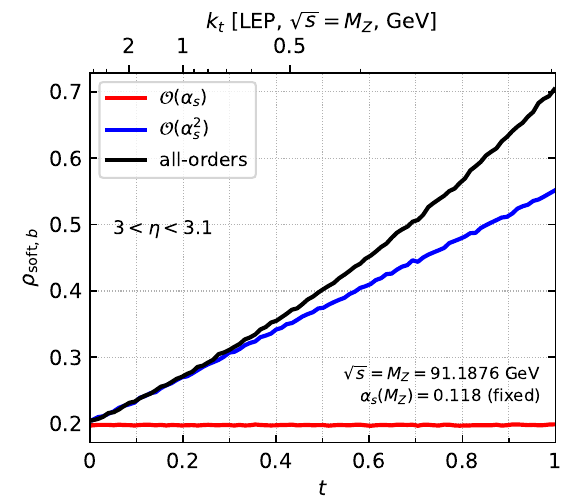}\hfill
\includegraphics[width=0.49\linewidth,page=6]{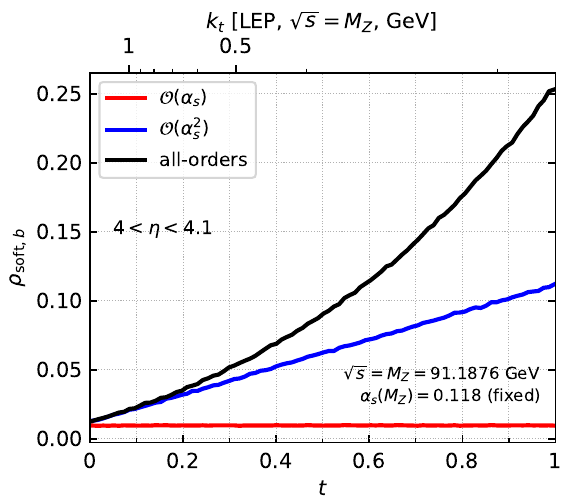}\hfill
    \caption{Soft contribution to the primary density for an initial
      $b \bar b$ dipole (at LEP energies), as a function of the
      evolution time, for two different slices in pseudorapidity
      $\eta$. In each plot, we show the all-order result (in black),
      and its $\order{\as}$ (red) and $\order{\as^2}$ (blue) expansions.}
    \label{fig:rho-soft}
\end{figure}
Numerical results for the soft resummation in the case of an initial $b \bar b$ dipole produced at LEP energies $\sqrt{s}=M_Z$ are shown in Figs.~\ref{fig: rho abelian and non abelian} and~\ref{fig:rho-soft}. 
In particular, the right-hand panel of Figs.~\ref{fig: rho abelian and non abelian}, shows the soft contribution to the primary density $\rho_b^{(\text{soft})}$ as a function of the angular value, here expresses in terms of the pseudorapidity of the emission, see Eqs.~(\ref{eq:lund-kinematic-variables-ee}), for a given slice in $z$. Here, large $\eta$ means small $\Delta$. Together with the all-order resummation, we also show its first and second order expansions, also obtained with our numerical implementation.  In order to check the implementation of our dipole shower, we compare the $\order{\as^2}$ term to the analytic calculation performed in the collinear limit, Eq.~(\ref{eq:final2gluons-HQ}) for $\mathcal{Q}=b$, finding excellent agreement. 
Finally, in Fig.~\ref{fig:rho-soft} we show $\rho_b^{(\text{soft})}$
as a function of the evolution time, for two different slices in
pseudorapidity $\eta$. In each plot, we show the all-order result (in
black), as well as its $\order{\as}$ and $\order{\as^2}$ expansions,
in red and blue, respectively.
The $x$-axis at the top of the plots show the corresponding $k_t$
scales at LEP energies.
At first order, the density has no logarithms, thus leading to a flat
distribution.
Additionally, the single-logarithmic behaviour of the next order is clearly visible.

 \section{Towards phenomenology}\label{sec:pheno}

In this section we combine all the effects discussed in this paper and provide all-order results for the Lund plane density. 
In order to present numerical results, we consider a simple process, namely $b \bar b$ production in $e^+e^-$ collisions at $\sqrt{s}=M_Z$. We discuss the impact of running coupling, collinear and soft resummation on the final result. We also perform matching to the exact $\order{\as}$ matrix element, with full $b$-mass dependence. Finally, we compare our findings to Monte Carlo simulations. 

\subsection{Resummed master formula}\label{sec:resummation-full}
\begin{figure}
\begin{minipage}{0.50\linewidth}
\includegraphics[width=\linewidth,page=2]{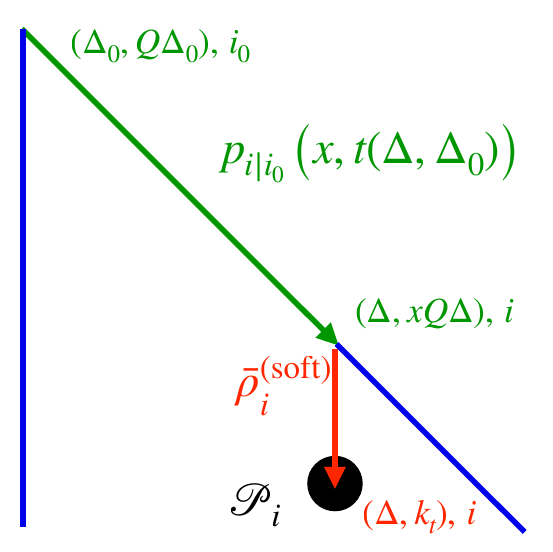}    \hfill
\end{minipage}
\begin{minipage}{0.50\linewidth}
    \caption{
    Pictorial representations of the different contributions that
    enter the resummation of the primary Lund plane density. The
    collinear evolution of the originating parton $i_0$ from the
    initial angular scale $\Delta_0$ to parton $i$ at an angle $\Delta$ is represented by a green arrow. During this evolution the transverse momentum is reduced from $Q \Delta_0$ to $x Q \Delta $. The red arrow represents instead soft evolution, at fixed angular scale, from $z Q \Delta$ to $\kt$. Finally, the black blob represents the splitting that is recorded.}
    \label{fig:cartoon}
    \end{minipage}
\end{figure}
First, let us assemble the various parts of our calculation and construct a resummed formula for the primary Lund plane density. 
We start from a hard parton of flavour $i_0$ at $(\Delta_0,Q\Delta_0)$, we use collinear evolution to obtain parton $i$ at the scale $(\Delta, xQ \Delta)$. This is pictorially represented by the green arrow in Fig.~\ref{fig:cartoon}.
At this point we use soft resummation to evolve from $xQ \Delta \to k_t$, at fixed $\Delta$, as shown by the red arrow. Finally, the splitting $\mathcal{P}_i$ (indicated by the black blob) is recorded.
In formulae, this reads
\begin{align}\label{eq:final-resummation}
  \rho_{i_0}^{(\text{res})}(\Delta,\kt)= \sum_{i=q,\mathcal{Q},g}\frac{\as^{\text{CMW}}(x_K^2 k_t^2)}{\pi}\int^1_0& \de x \, p_{i|i_0}(x, t(\Delta,\Delta_0))\, z \, \mathcal{P}_i(1-z)\Theta(1-2z)\Big|_{z= \frac{k_t }{x Q \Delta}} \nonumber\\ 
  &\cdot {\bar \rho}^{(\text{soft})}_{i} \left(\Delta, \tilde{z}_\text{eff}(x)\, Q \Delta\right)
  ,
\end{align}
where the bar indicates that we have divided out the small-angle limit at $\order{\as}$ in order to avoid double counting with $\mathcal{P}_j$, see Eqs.~$(\ref{eq:final2gluons-massless})$, $(\ref{eq:final2gluons-gluon})$, and~(\ref{eq:final2gluons-HQ}).
We note that for $i=\mathcal{Q}$ in the above expression, the
collinear evolution freezes at $\Delta=\Delta_d$. This is coherent
with our picture for the resummation of multiple soft emissions in
$\rho^\text{(soft)}$ where the dominant contribution comes from
multiple emissions at an angular scale commensurate with $\Delta_d$.

Following~\cite{Lifson:2020gua}, we have also introduced
\begin{equation}
 \tilde{z}_\text{eff}(x)= \exp \left[ -\frac{\pi}{\as}t\left(\frac{x_Z \,x \,\kt}{x-(2-x_Z)\frac{\kt}{Q \Delta}}, x Q \Delta\right)\right],
\end{equation}
so that $t\to 0$ (and hence $z_\text{eff} \to 1$) when $\kt \to \frac{1}{2} xQ \Delta$.
As already done in Sec.~\ref{sec:soft-small-angle-as}, we have attached a flavour label to the soft density. While this is well-defined in the small-angle approximation, this assignment is ambiguous for the full $\rho^{(\text{soft})}$. 
However, we can still combine together all the dipole configuration that contribute to a given jet  and thus separate $\rho^{(\text{soft})}$ according to the flavour label, $i =q,\mathcal{Q}, g$.
There is an additional subtlety. In Eq.~(\ref{eq:final-resummation}), soft evolution starts at the scale $x Q \Delta$ for a parton of flavour $i$ that needs not to be present at Born level. Thus, we also need to generate the evolution for all possible QCD dipoles, not just the one present in the Born process.
In particular, one sees that at large angles,
where the details of the dipole configuration matter, collinear flavour changing eﬀects can be neglected and only the dipole configurations present at Born level contribute. In the opposite regime, i.e.\ at small
angles, where collinear evolution matters, the flavour assignment to $\rho_i^{(\text{soft})}$ is unambiguous. 

In order to assess the theoretical uncertainty related to our calculation, we consider different types of scale variation. First, we vary renormalisation scale $\mur$, which is the reference scale for the strong coupling, by a factor of 2 around the hard scale $Q$. Then, following~\cite{Lifson:2020gua},  we also vary in Eq.~(\ref{eq:final-resummation}) the factor $x_K$ between $\frac{1}{2}$ and $2$ in order to probe running-coupling corrections and the factor $x_Z$ to probe soft effects. Finally, the uncertainty on the collinear resummation is estimated by varying the starting angular scale $\Delta_0$.

\subsection{The Lund $b$-jet plane in $e^+e^-$ collisions}
\begin{figure}[h!]
\begin{minipage}[b]{0.50\linewidth}
\includegraphics[width=0.98\linewidth,page=7]{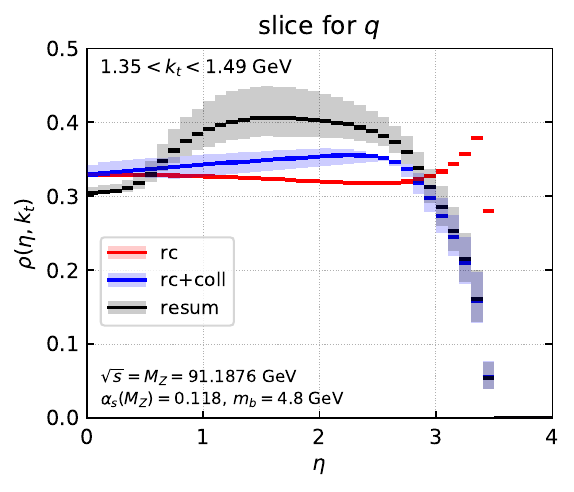}\hfill
\end{minipage}\hfill
\begin{minipage}[b]{0.45\linewidth}
  \caption{Resummed Lund plane density in $e^+e^-$ collisions at the $\sqrt{s}=M_Z$, as a function of the pseudorapidity $\eta$, for a representative slice in $\kt$.
  The top plot is for hemisphere jets initiated by a light quark
  ($q$), the ones in the middle are for $b$-jets, with standard
  declustering (left) or following the flavoured branch (right), and the plots at the bottom show the ratio $b/q$. 
  Each plot features three curves: running coupling resummation (red), running coupling and collinear resummation (blue), and the full resummation (black).} \label{fig:resum_1D}
\end{minipage}
\includegraphics[width=0.49\linewidth,page=8]{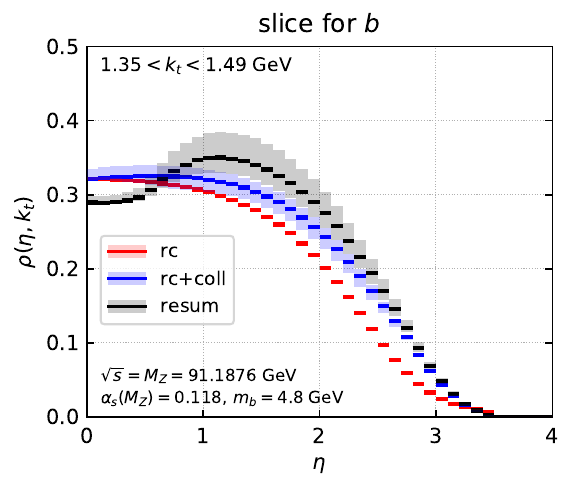}\hfill
\includegraphics[width=0.49\linewidth,page=9]{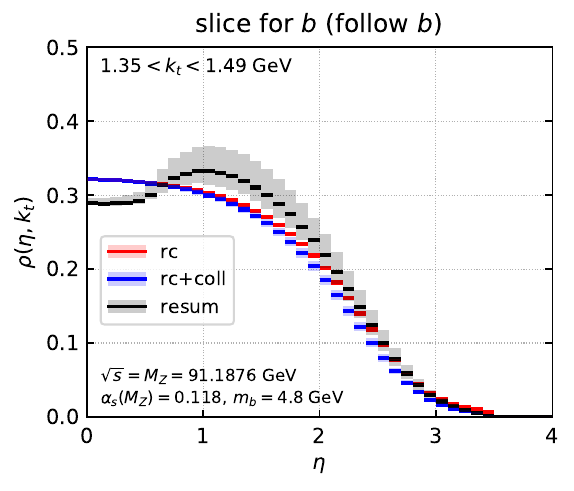}
\includegraphics[width=0.49\linewidth,page=15]{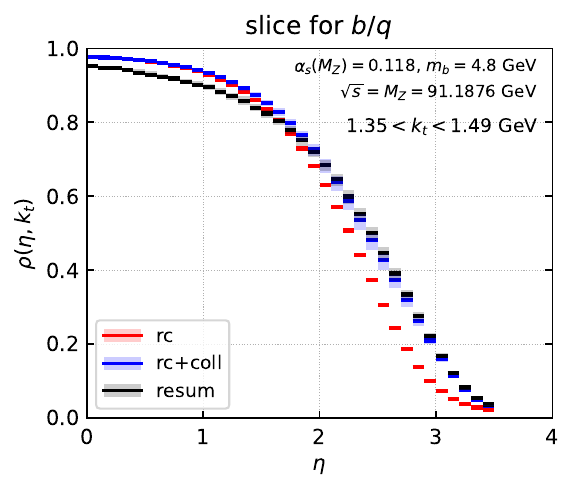}\hfill
\includegraphics[width=0.49\linewidth,page=16]{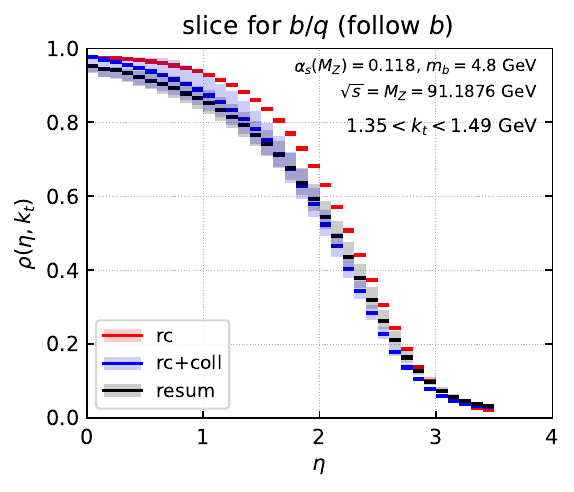}
\end{figure}
We are now ready to present our numerical results. For this purpose, we focus on $b \bar b$ production in $e^+e^-$ collisions at $\sqrt{s}=M_Z$, as well as light-quark production. Specifically, we partition each event into two hemisphere jets, defined with the C/A algorithm, and measure the primary Lund plane density on both. 
Note that at large $k_t$ the primary Lund plane density has a non-zero
contribution for $\eta<0$ which neither shown nor discussed here.
We present resummed results for the density expressed in terms of pseudorapidity and transverse momentum, namely $\widetilde{\rho}(\eta, \kt)$ defined in Eq.~(\ref{eq:density-def-eta-kt}).
However, to avoid cumbersome notation, in this section, we rename 
$\widetilde{\rho}(\eta, \kt) \to \rho(\eta, \kt)$.
Note that with the choice of kinematic variables $(\eta,\kt)$, the most natural scale for the reference hard scale is the collision energy rather than the jet energy, i.e.\ $Q=\sqrt{s}=2 E_\text{jet} $, essentially because the pseudorapidity is defined in terms of splitting angles divided by two, see Eq.~(\ref{eq:lund-kinematic-variables-ee}).

We start by considering the size of the different contributions to the all-order results that we have discussed in this paper. To this purpose, we consider three approximations:
\begin{enumerate}
    \item[a)] rc: running coupling resummation is considered, but both collinear and soft resummation are neglected, i.e. $p_i\propto \delta(1-x)$ and $\rho^{(\text{soft})}$ is computed at $\mathcal{O}(\as)$;
    \item[b)] rc+coll: collinear evolution is turned on but
      $\rho^{(\text{soft})}$ is still computed at $\mathcal{O}(\as)$
      (i.e.\ $\bar\rho^\text{(soft)}$ is computed at order $\alpha_s^0$);
    \item[c)] resum: the full rc+coll+soft resummation is considered;
\end{enumerate}
The numerical size of these effects is shown in Fig.~\ref{fig:resum_1D}.  In particular, each plot shows a representative $\kt$ slice of the Lund density, as a function of the pseudorapidity $\eta$.
The top plot is for jets initiated by a light quark ($q$), while the ones in the middle are for $b$-jets, with standard declustering (on the left) or following the flavoured branch (on the right).
In particular, the three approximation to the resummed results discussed above are reported. 
First, we note that, when we use standard declustering, running coupling resummation (shown in red) receives substantial corrections at large rapidities, i.e.\ at small angles, when the collinear resummation is turned on (shown in blue). This is likely due to the matrix structure of collinear evolution that allows for flavour mixing. Indeed, this effect is very much reduced when we use flavour declustering. In this case, we recover the collinear evolution of the flavour non-singlet contribution, which is diagonal. 
We note that the full resummation, including soft effects (shown in black) closely follows the collinear at large rapidities. However, we do see sizeable effects at moderate $\eta$. This is easy to understand by looking at our cartoon in Fig.~\ref{fig:cartoon}: at intermediate rapidities, we probe a wider evolution in $\kt$ (therefore soft effects can be more pronounced). 
The plots at the bottom of Fig.~\ref{fig:resum_1D} represent instead
the ratio $b/q$. First, we note that this ratio is close to unity at small $\eta$, but it decreases at larger rapidities. This is a clear manifestation of the dead cone. Further, we note that by taking this ratio, the theoretical uncertainty, as measured by varying the perturbative scales of our resummed results, as described in the previous sections, is very much reduced. 

\begin{figure}
\begin{minipage}{0.50\linewidth}
\includegraphics[width=0.98\linewidth,page=11]{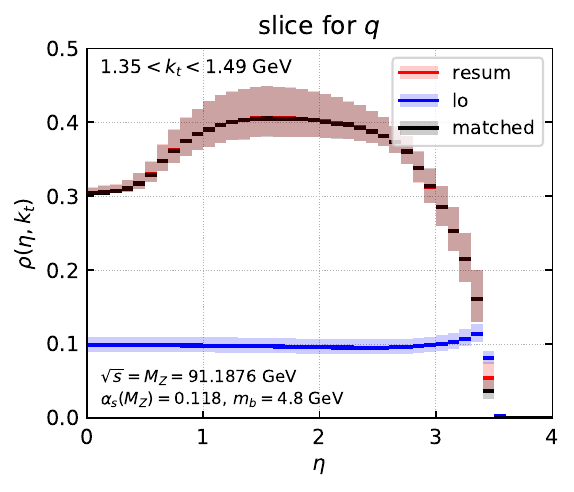}\hfill
\end{minipage}\hfill
\begin{minipage}[b]{0.45\linewidth}
  \caption{Same as in Fig.~
\ref{fig:resum_1D} but now each panel shows: resummed result (red), fixed-order at tree level (blue) and matched result (black), obtained with Eq.~(\ref{eq:matched_density}).}    \label{fig:resum_matched_1D}
\end{minipage}
\includegraphics[width=0.49\linewidth,page=12]{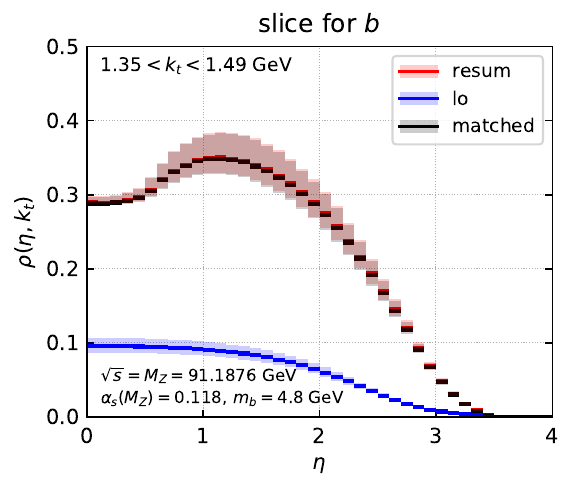}\hfill
\includegraphics[width=0.49\linewidth,page=13]{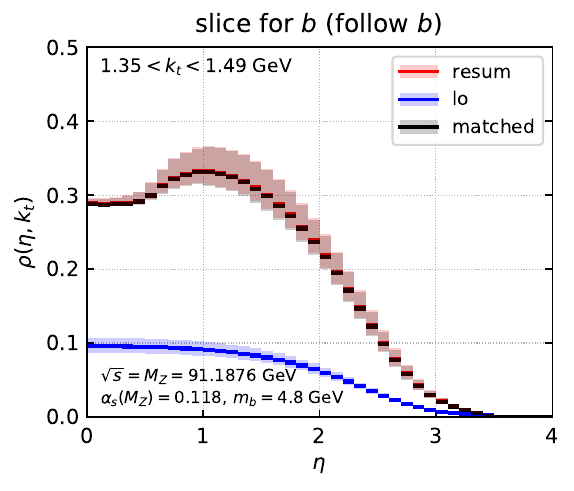}
\includegraphics[width=0.49\linewidth,page=17]{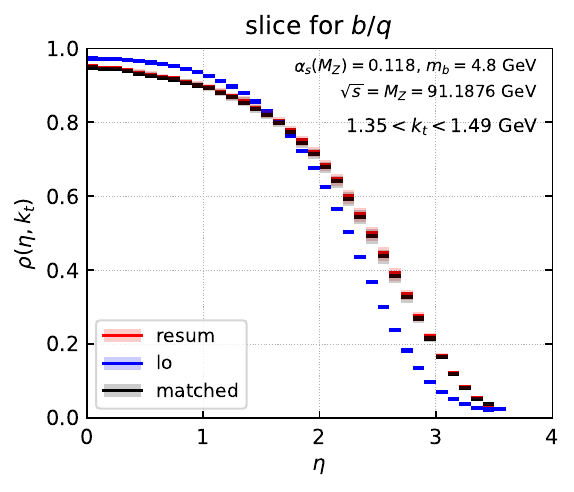}\hfill
\includegraphics[width=0.49\linewidth,page=18]{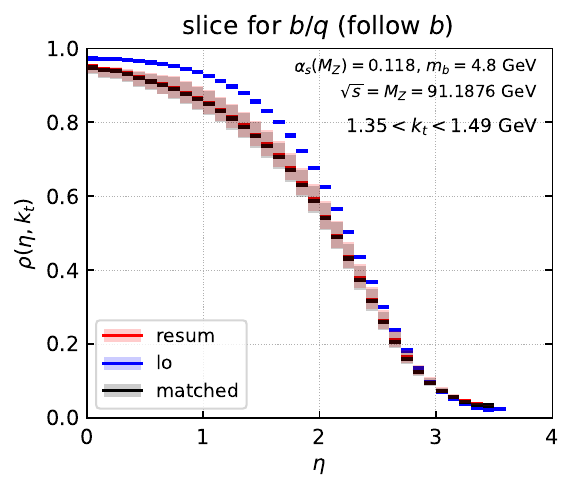}
\end{figure}

Next, we perform matching of our all-order result to the Lund plane density computed at $\order{\as}$.
In particular, the matched Lund plane density is given by
\begin{align}\label{eq:matched_density}
\rho_{i_0}^{(\text{matched})}(\eta,\kt)=  \rho_{i_0}^{(\text{res})}(\eta,\kt)\,\frac{\rho_{i_0}^{(\text{f.o.})}(\eta,\kt)}{\rho_{i_0}^{(\text{res},\as)}(\eta,\kt)},\quad i_0=b,q,
\end{align}
where in the denominator we have the expansion of the resummed result to first order in the strong coupling. 
Note that, with a slight abuse of notation, we have used the index $i_0$ also in the fixed-order expression to denote whether the $\order \as$ calculation has been done with massive or massless quarks. 
Details are given in App.~\ref{app:e+e-lo}. 
Our findings are shown in Fig.~\ref{fig:resum_matched_1D}. The figure plots follow the same pattern as in Fig.~\ref{fig:resum_matched_1D}, but now fully resummed results are shown in red, fixed-order ones, computed at leading order, are shown in blue and the matched ones in black. 
We note a rather striking difference between the fixed-order and the resummation, both for light and massive quarks. However, this large difference is largely due to the fact that the fixed-order calculation is done at fixed coupling $\as(M_Z)$ while in the resummed one we use $\as(\kt)$. Indeed this effect cancels to a large extend in the matching and the final result is rather close to the resummation.
Therefore, the conclusions reached for the resummed results also hold for the matched ones. 
    
\begin{figure}
\centering
\includegraphics[width=0.47\linewidth,page=2]{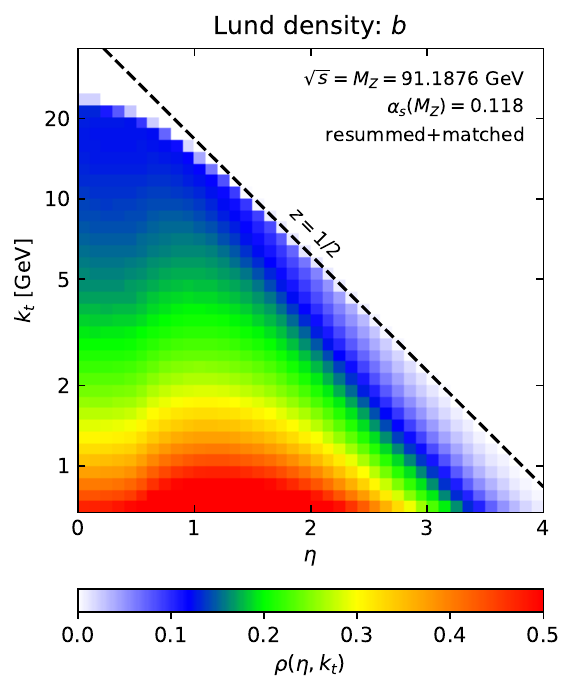}\hfill
\includegraphics[width=0.47\linewidth,page=3]{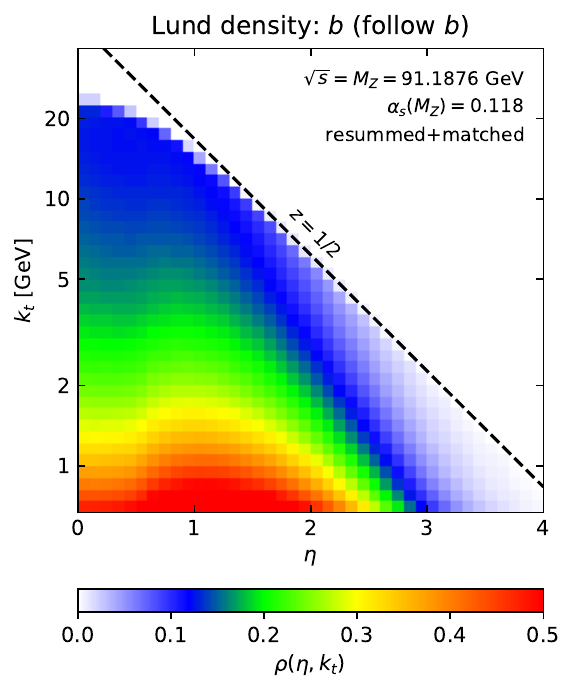}
\includegraphics[width=0.47\linewidth,page=5]{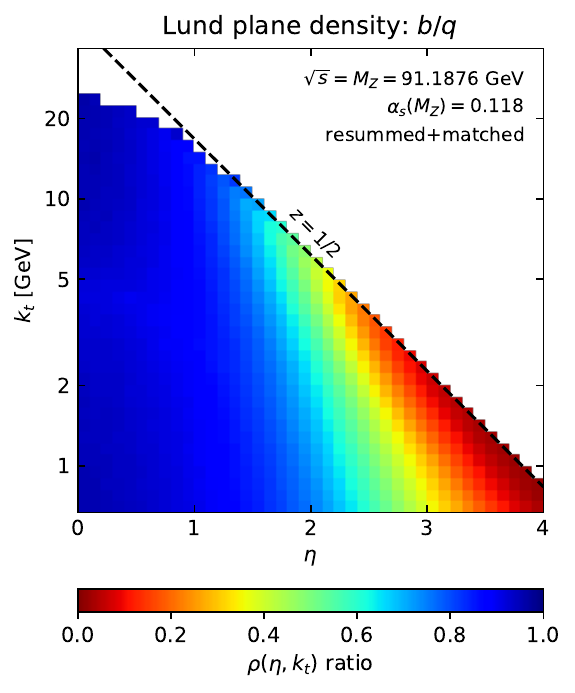}\hfill
\includegraphics[width=0.47\linewidth,page=6]{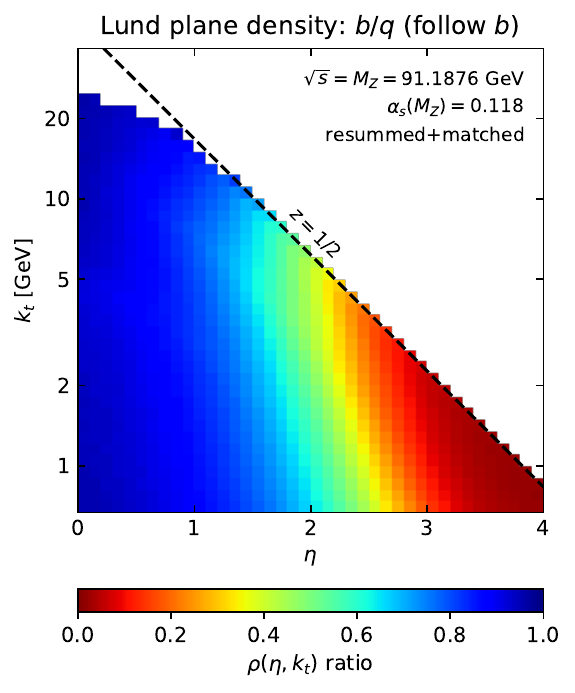}
    \caption{Resummed and matched results for the primary Lund plane density in $e^+e^-$ collisions at the $\sqrt{s}=M_Z$. The top plots are for hemisphere jets initiated by a $b$-jets, with standard declustering (on the left) or following the flavoured branch (on the right). The plots at the bottom show the ratio $b$-jet density over the density for a jet initiated by a massless quark $q$. The dashed line corresponds to $z=1/2$, which represents the phase-space boundary.}
    \label{fig:resum_matched_2D}
\end{figure}

We conclude this section by showing the resummed and matched results for the two dimensional Lund plane density in Fig.~\ref{fig:resum_matched_2D}. The plots at the top are the Lund $b$-jet plane density with the two declustering schemes used in this work, while the plots at the bottom represent the ratio of the $b$-jet densities in the two cases to the light-quark density (not shown here). The ratio plots clearly show a depletion of radiation at large rapidity, which is indeed a direct manifestation of the dead cone. Note that the effect is visible also in the perturbative region, i.e.\ above $\kt \sim 1$~GeV. Dead-cone effects are enhanced when the flavour-declustering procedure is used. This is expected because this avoids flavour mixing in the collinear evolution.

Finally, although this study is for $e^+e^-$ collisions, it can teach us some lessons also for the LHC as well. In hadron-hadron collision, the hard scale is provided by the transverse momentum of the jet times the jet radius. Thus, the $e^+e^-$ results explored in this section roughly corresponds to jets of $p_t R_0 \simeq E_\text{jet}= \frac{M_Z}{2}$ in $pp$ collisions. For $R_0=0.4$, this gives $p_t\simeq 110$~GeV. However, this is only a qualitative picture because further effects need to be accounted for in $pp$ collisions, both at perturbative level, e.g.\ different partons initiating the collinear evolution and a more dipoles contributing to soft resummation, as well as non-perturbative contributions such as the Underlying Event. 

\FloatBarrier

\subsection{Comparison to Monte Carlo}\label{sec:MC}
\begin{figure}
    \centering
    \includegraphics[width=0.33\linewidth,page=1]{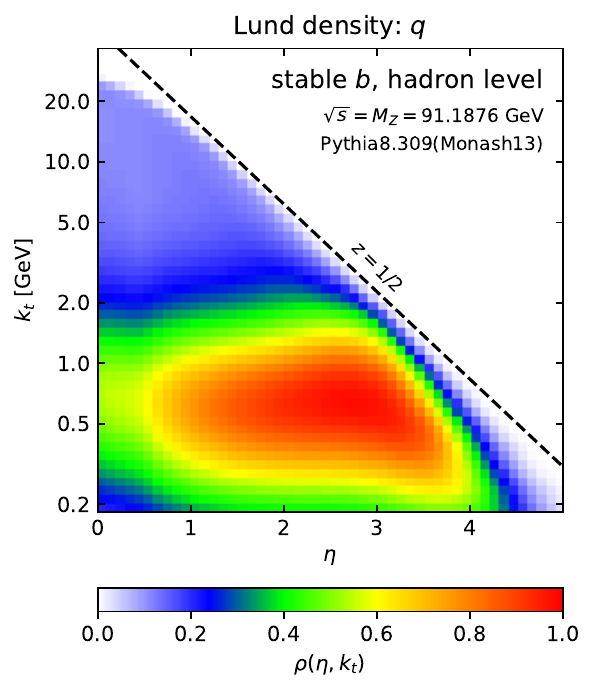}\hfill
    \includegraphics[width=0.33\linewidth,page=2]{figures/pythia8.pdf}\hfill
    \includegraphics[width=0.33\linewidth,page=5]{figures/pythia8.pdf}
    \includegraphics[width=0.33\linewidth,page=4]{figures/pythia8.pdf}\hfill
    \includegraphics[width=0.33\linewidth,page=3]{figures/pythia8.pdf}\hfill
    \includegraphics[width=0.33\linewidth,page=6]{figures/pythia8.pdf}
    \caption{Monte Carlo simulation of the primary Lund plane density for hemisphere jets in $e^+e^-$ collisions at $\sqrt{s}=M_Z$. The plots on the left and at the centre show the primary Lund plane density for light jets (top-left), $b$-jets with standard declustering (top-centre), $b$-jets with follow-the-flavour declustering (bottom-right), all obtained with stable $B$ hadrons. The bottom-left plot instead shows $b$-jets density with $B$-hadron decays.
    The plots on the right instead show the ratios $b/q$.
    }
    \label{fig:mc-1}
\end{figure}

We conclude our work by performing a study using a general-purpose Monte Carlo event generator. This allows us to test our analytic predictions against pseudo-data and to assess the size of non-perturbative corrections. 
For this study we simulate $e^+e^-$ collisions at $\sqrt{s}=91.1876$~GeV, using \pythia{8.309}~\cite{Bierlich:2022pfr}, with the Monash tune~\cite{Skands:2014pea}. We switch off QED radiation, as well as $\pi^0$ and $B$ decay.
The primary Lund plane density is measured on each of the two hemisphere jets defined with the exclusive C/A algorithm. Results are presented in Fig.~\ref{fig:mc-1}, where we show the densities for light jets and for $b$-jets using the two declustering approaches discussed in this work, namely the standard one and the one following the $b$-branch.
The suppression of radiation at large rapidities, i.e.\ at small angles, is clearly visible and it is numerically demonstrated by the $b/q$ plots. The results obtained with Monte Carlo simulations exhibit a behaviour that is comparable, at least qualitatively, to what we have obtained with our analytic studies. 

Before turning to a more quantitative comparison, let us briefly discuss the role of $B$ decays.
This is done in bottom-left plot in Fig.~\ref{fig:mc-1}. For this simulation, we turn $B$-decays on. 
As already discussed in our previous study~\cite{Dhani:2024gtx},
$B$-hadron decays considerably alter the  substructure of
heavy-flavour jets. Therefore, unless one is interested in studying
the properties of the decay products, our recommendation for
heavy-flavour jet substructure measurements is to always reconstruct
the $B$-hadron kinematics before proceeding with the jet substructure
analysis. 

\begin{figure}
    \centering
    \includegraphics[width=0.49\linewidth,page=12]{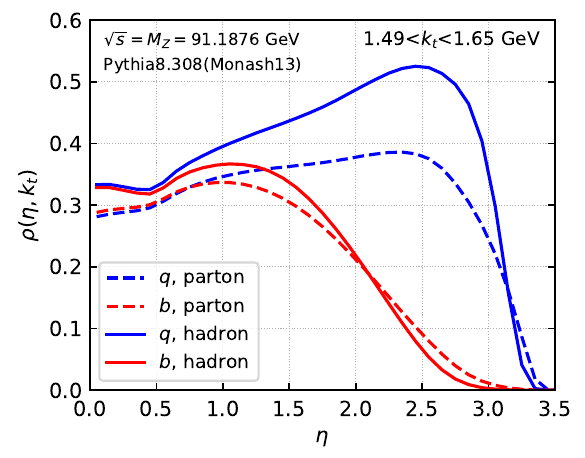}\hfill
    \includegraphics[width=0.49\linewidth,page=13]{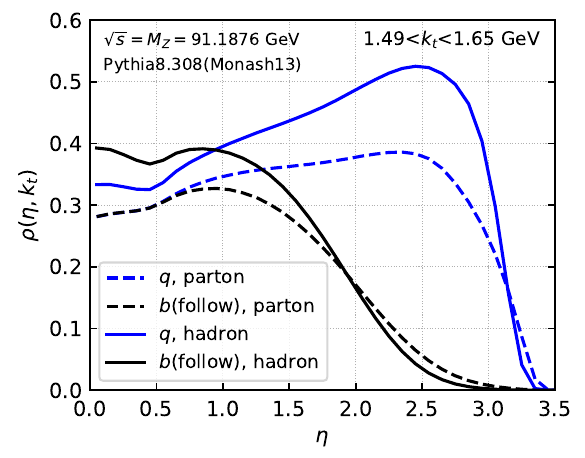}
    \includegraphics[width=0.49\linewidth,page=14]{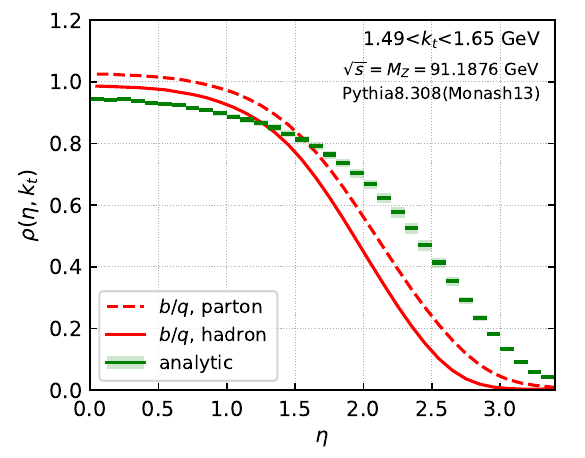}\hfill
    \includegraphics[width=0.49\linewidth,page=15]{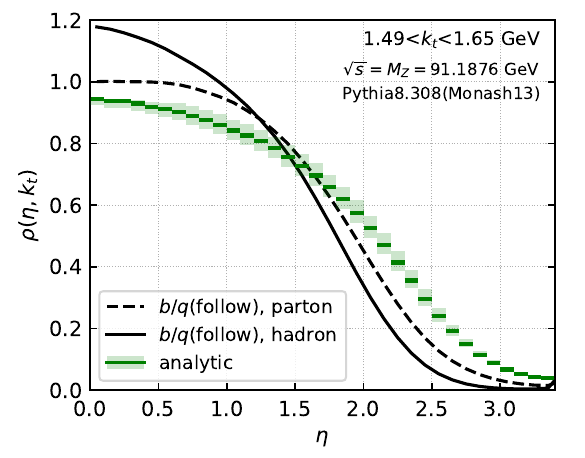}
    \caption{Slice of the Lund plane density at fixed $\kt$, as a function of $\eta$. Results are obtained with Monte Carlo simulations and compared to the resummed and matched calculation.
     }
    \label{fig:mc-2}
\end{figure}

In order to perform a more quantitative comparison of the Monte Carlo
simulation and our resummed and matched result, we consider a slice in
$\kt$ and we plot the density obtained with the Monte Carlo
simulation, as a function of $\eta$. This is done in Fig.~\ref{fig:mc-2}. In both plots at the top, the blue curves are for the case of light jets: dashed for parton-level and solid for hadron-level. The top plot on the left then shows the case of a $b$-jet (in red) obtained with standard declustering, while the one on the right is done with the follow-the-flavour declustering (in black). 
We first note, that while the latter has a simpler analytic structure that avoids light-parton contamination in the dead-cone region, it is also more sensitive to hadronisation corrections. 

It is then interesting to consider the ratios $b/q$ and compare the results obtained with the parton shower to our resummed and matched prediction.
We do this in the bottom plots of Fig.~\ref{fig:mc-2}, on the left for the standard declustering, on the right for the follow-the-$b$ procedure. The first interesting thing to note is that non-perturbative hadronisation corrections largely cancel when looking at $b/q$, making the ratio a good observable to study perturbative dead-cone effects.~\footnote{We should note that we are considering here only quark jets, as appropriate for $e^+e^-$ collisions. At the LHC this cancellation might be less pronounced because of the differences between quark- and gluon-jets.}
However, we also note that our resummed and matched results exhibit some differences with respect the Monte Carlo simulation. The agreement is not bad for flavoured branch procedure, while we observe a significant discrepancy for the standard declustering procedure. In particular, our calculations always predict a higher-ratio, i.e.\ less dead-cone suppression, than the Monte Carlo.

The poor agreement we observe in the case of standard declustering is somewhat puzzling.
In our approach the effect of collinear and soft resummation is to fill the dead cone, as can be seen in bottom-left plot of Fig.~\ref{fig:resum_1D}, where the running-coupling-only curve is below the other ones. 
Thus, it is surprising that the parton-shower result is closer to the naive running-coupling result than to the one which includes the resummation of collinear and soft logarithms. 
    
\section{Conclusions and Outlook}\label{sec: conclusion}

In this paper we have studied the primary Lund plane density for jets initiated by a heavy quark. 
We have performed a detailed analysis of all the kinematic regions that are relevant to single-logarithmic accuracy. In particular, the presence of the heavy-quark introduce a further scale in the calculation and therefore introduces further complications with respect to the light-parton case, studied in~\cite{Lifson:2020gua}. 
Although, we have presented explicit results for the case of $b$-jets, our findings can be straightforwardly extended to $c$-jets too.

At single-logarithmic accuracy, three effects must be considered, namely running coupling corrections, enhanced by logarithms of the transverse momentum \(k_t\); hard-collinear logarithms of the emission angle \(\Delta\), which induce flavour-changing effects; soft emissions at large and commensurate angles, enhanced by logarithms of $\frac{\kt}{Q\Delta}$, including the intricate structure of clustering logarithms. 
In all cases, the presence of the heavy-flavour introduces complications that we have addressed by adopting a variable flavour-number scheme for both the running of the strong coupling and the collinear evolution. Furthermore, we have accounted for finite-mass effects in the resummation of soft logarithms, which is achieved, numerically, in the large-$\nc$ limit, by means of a $\kt$-ordered dipole shower.  Finally, we have employed the quasi-collinear limit, with massive splitting functions, to describe the splitting that is recorded.

The main scope of this paper is to provide solid theoretical foundations to the use of the Lund plane density to study the dead cone, namely the suppression of radiation within a cone of radius $\Delta_d=m/Q$ along the heavy-quark direction. 
In this context, we have found that the naive picture, obtained in the soft and quasi-collinear limit, of a suppression $\left(\frac{\Delta^2}{\Delta^2+\Delta_d^2}\right)^2$, receives multiple corrections. First, collinear evolution can change the flavour of the leading parton, resulting in radiation down to angular scales smaller than the dead cone. The bigger the hard scale, e.g.\ the jet transverse momentum, the larger this effect. In such cases, in order to maintain sensitivity to the dead-cone, one can change the declustering procedure, thereby always following the flavoured branch~\cite{Cunqueiro:2018jbh}. We have computed resummed results for both declustering procedures. 
Second, we have uncovered a rather surprising effect that originates
from correlated soft-gluon emissions at commensurate angles, for which
the dead-cone acts as a boundary, giving rise to contributions
to the Lund plane density deep inside the dead cone dominated by
multiple gluon emissions at much larger angles commensurate with the
dead cone angle. In this context, we have explicitly computed the $\order{\as^2}$ contribution and shown that it produces a dead-cone suppression which is surprisingly milder than the one-gluon case. 
We have developed a numerical code that takes into account all these effects, thus allowing us to perform single-logarithmic resummation for the density of the primary Lund $b$-jet plane. The resummation codes can be interfaced with fixed-order matrix elements generators to produced resummed and matched results for $b$-jet production for processes in hadron or lepton collisions.

Although the focus of this paper is to highlight and address all the novel ingredients that enter the resummation of the Lund $b$-jet plane at single-logarithmic accuracy, we have started moving the first steps towards phenomenology. To this purpose we have considered the simplest possible (but interesting) scenario, namely hemisphere jets in $e^+e^-$ collisions at $\sqrt{s}=M_Z$. We have compared our resummed and matched prediction to results obtained with Monte Carlo event generators. 
We have found qualitative agreement between the two approaches, although a more detailed analyses reveals that the resummed and matched calculation predicts smaller suppression in the dead-cone region than what is obtained with a standard parton shower. 
The agreement between the resummation and Monte Carlo predictions improve if the the flavoured-branch declustering procedure is adopted. 

We see this paper as a first necessary step towards phenomenological investigation of the fragmentation properties of heavy-quark jets exploiting the Lund plane. We plan to continue pursuing this research line by computing theoretical predictions that can be directly compared to existing or upcoming data from the LHC collaborations. 
Further, we believe that our work will prove useful in the context of including quark-mass effects in high-accuracy parton shower algorithms~\cite{Assi:2023rbu,Hoche:2024dee}. 
Finally, we believe our formalism can be extended to the examine Lund planes of multi-pronged jets, such as those originated from electroweak bosons ($Z/W/H$), $g \to b\bar b$, and top quarks and we look forward to exploring these research avenues.

\paragraph{Acknowledgements.} 
The work of A.G. and S.M. was supported by the Italian Ministry of Research (MUR) under grant PRIN 2022SNA23K funded by the European Union -- Next Generation EU, Mission 4, Component 2, CUP D53D23002880006 and by ICSC Spoke~2 under grant BOODINI. 
The work of A.G. is also supported by the Excellence Cluster ORIGINS, funded
by the Deutsche Forschungsgemeinschaft (DFG, German Research Foundation) under Germany’s
Excellence Strategy — EXC-2094-390783311 and by a TUM Global Postdoc Fellowship.
A.G. and S.M. wish to thank the EIC Theory Institute at Brookhaven National Laboratory, for hospitality and support during the course of this work.

\FloatBarrier
\appendix 

\section{Collinear evolution and Mellin moments}\label{app:mellin}
In this appendix we report details of the (quasi)-collinear calculations.
We start by introducing the unregularised splitting functions:~\footnote{Note that the variable $z$ in this appendix should be taken as just the argument of the splitting functions. It is not the Lund plane variable.} 
\begin{alignat}{2}
P_{qq}(z) &= \cf \left( \frac{1 + z^2}{1 - z} \right), 
&P_{gq}(z) &= P_{qq}(1 - z), \nonumber\\
P_{\mathcal{QQ}}(z) &= \cf \left( \frac{1 + z^2}{1 - z} - \frac{2 m^2 z(1 - z)}{q_t^2 + (1 - z)^2 m^2} \right), 
&\quad P_{g\mathcal{Q}}(z) &= P_{\mathcal{QQ}}(1 - z), \nonumber\\
P_{gg}(z) &= 2\ca \left( \frac{z}{1 - z} + \frac{1 - z}{z} + z(1 - z) \right), 
&\quad P_{\mathcal{Q}q}(z) &= P_{q\mathcal{Q}}(z) =0 , \nonumber\\
\quad P_{qg}(z) &= \tr \left( z^2 + (1 - z)^2 \right), \quad& P_{\mathcal{Q}g}(z) &= \tr \left( 1 - \frac{2 z(1 - z) q_t^2}{m_h^2 + q_t^2} \right),
\end{alignat}
where $\cf=\frac{4}{3}, \, \ca=3, \, \tr=\frac{1}{2}$ are the standard colour factors and Gellman matrices normalisation, respectively.

The splitting kernels $P^{(R)}$ and  $P^{(V)}$ introduced in  Eq.~\eqref{eq: DGLAP type} read :
\begin{subequations}
\begin{align}
     P_{qq}^{(R)}(z) & = P_{qq}(z)\, \Theta(2z-1), &
    P_{qq}^{(V)}(z) & = P_{qq}(z), \\
    P_{gq}^{(R)}(z) & = P_{gq}(z)\,\Theta(2z-1),  &
   P_{gq}^{(V)}(z) & = 0, \\
      P_{gg}^{(R)}(z) & = P_{gg}(z)\,\Theta(2z-1), &
   P_{gg}^{(V)}(z) & = \frac{1}{2}P_{gg}(z) + n_f P_{qg}(z), \\
    P_{qg}^{(R)}(z) & = 2 (n_f-1)P_{qg}(z)\,\Theta(2z-1), &
     {P}_{qg}^{(V)}(z) & = 0, \\
    {P}_{\mathcal{Q}\mathcal{Q}}^{(R)}(z) & = P_{qq}(z)\, \Theta(2z-1), &
    {P}_{\mathcal{Q}\mathcal{Q}}^{(V)}(z) & = P_{qq}(z), \\
    {P}_{\mathcal{Q}q}^{(R)}(z) & = 0, &
    {P}_{\mathcal{Q}q}^{(V)}(z) & = 0, \\
    {{P}}_{q\mathcal{Q}}^{(R)}(z) & = 0, &
    { {P}}_{q\mathcal{Q}}^{(V)}(z) & = 0, \\
    { {P}}_{g\mathcal{Q}}^{(R)}(z) & = P_{gq}(z)\, \Theta(2z-1),  &
    P_{g\mathcal{Q}}^{(V)}(z) & = 0, \\
  { {P}}_{\mathcal{Q}g}^{(R)}(z) & = 2 P_{q g}(z)\, \Theta(2z-1), & \quad  { {P}}_{\mathcal{Q}g}^{(V)}(z) & = 0.
\end{align}
\end{subequations}
Given the expressions of $P^{(R)}$ and $P^{(V)}$ we notice that Eq.~(\ref{eq: DGLAP type}) can be also be expressed as:
\begin{align}
\label{eq: modified dglap}
    \frac{\de}{\de t} p_i(x,t)= \sum_{j=l,h,g} \int^1_x \frac{\de z}{z} \left(\hat{P}_{ij}(z)-\delta \hat{P}_{ij}(z)\right) p_{j} \left(\frac{x}{z},t\right),
\end{align}
where $\hat{P}_{ij}(z)$ are the regularised massless splitting functions and $\delta\hat{P}_{ij}(z)= \hat{P}_{ij}\, \Theta(1-2z)$:
  \begin{align}
  \label{eq: reg splitting functions}
  \begin{array}{ll}
    \hat{P}_{qq}(z) = \hat{P}_{\mathcal{Q}\mathcal{Q}}(z) = \cf\left(\frac{1+z^2}{1-z}\right)_+, &
    \hat{P}_{gq}(z) = \hat{P}_{g\mathcal{Q}}(z) = P_{gq}(z), \\[1ex]
    \hat{P}_{gg}(z) = 2 \ca\left[\frac{z}{(1-z)_+}+\frac{1-z}{z}+z(1-z)\right]
    + 2\pi \beta_0^{(n_f)} \delta(1-z), &
    \hat{P}_{qg}(z) = \hat{P}_{\mathcal{Q}g}(z) = P_{qg}(z).
  \end{array}
\end{align}
We immediately notice that for $x>1/2$:
\begin{align}
    \int^1_x \frac{\de z}{z} \delta\hat{P}_{ij}(z) p_{j} \left(\frac{x}{z},t\right)= \int^1_x \frac{\de z}{z} \hat{P}_{ij}(z) p_{j} \left(\frac{x}{z},t\right) \Theta(1-2z)
=0.
\end{align}
Therefore, for $x>1/2$, Eq. (\ref{eq: modified dglap}) reduces to the standard DGLAP equation. 
Moreover, imposing the initial condition Eq.~\eqref{eq: initial cond}, we can formally write the solution of the differential equation as:
\begin{align}
\label{eq: formal solution}
    p_{i}(x,t)= \delta_{i i_0}\delta(1-x)+\sum^\infty_{n=1} \frac{t^n}{n!} \bigotimes^n_{i=1}\hat{P}(z_i) \Theta(2z_i-1),
\end{align}
where:
\begin{align}
\label{eq: convolution def.}
    \bigotimes^n_{i=1}\hat{P}(z_i) \Theta(2z_i-1)= \int^1_{\frac{1}{2}} \de z_1 \dots \int^1_{\frac{1}{2}} \de z_n \hat{P}_{i j_1}(z_1) \dots \hat{P}_{j_{n-1} i_0}(z_n) \delta(z_1 \dots z_n-x).
\end{align}
From Eq. (\ref{eq: convolution def.}), we observe that the $n^{\text{th}}$ term of the series in eq. (\ref{eq: formal solution}) has support only for $x>2^{-n}$.
\subsection{Mellin space solution}
We define the Mellin transform of a generic function $f(x)$ as:
\begin{align}
    \tilde{f}(N)=\int^1_0 \de x\, x^{N-1} f(x).
\end{align}
By taking the Mellin transform, of Eq.~(\ref{eq: modified dglap}), we find:
\begin{align}
 \frac{\de}{\de t}\tilde{p}_i(N,t)= \Gamma_{ij}(N)\; \tilde{p}_j(N,t),
\end{align}
where the the anomalous dimension $\Gamma_{ij}$ is obtained as:
\begin{align}
\Gamma_{ij}(N)&=\gamma_{ij}(N)-\delta \gamma_{ij}(N)= \int_0^1 dx\, x^{N-1} \hat{P}_{ik}(x) - \int_0^{1/2} dx\, x^{N-1} \hat{P}_{ij}(x)\,.
\end{align}
The standard moments $\gamma_{ij}$ of the splitting functions are \cite{Gross:1973zrg,Georgi:1974wnj,Altarelli:1977zs}
\begin{align}
\gamma_{qq}(N) &= \cf\left(
\frac{3}{2}+\frac{1}{N(N+1)} - 2H_N
\right)\,,\\
\gamma_{qg}(N)&=\tr\,
\frac{N^2+N+2}{N(N+1)(N+2)}\,,\\
\gamma_{gq}(N) &= \cf\, \frac{N^2+N+2}{N(N^2-1)}\,,\\
\gamma_{gg}(N)&=\ca\left(
\frac{2}{N(N-1)}+\frac{2}{(N+1)(N+2)}-2H_N
\right) +2\pi \beta_0^{(n_f)},
\end{align}
where $H_N$ is the harmonic number.  The modifications to these moments are the same as the ones derived for the WTA flavour evolution~\cite{Caletti:2022glq} and they read
\begin{align}
\delta\gamma_{qq}(N) &= \cf\left(
B_{1/2}(N,0)+B_{1/2}(N+2,0)
\right)\,,\\
\delta\gamma_{qg}(N)&=\tr\,
\frac{N^2+3N+4}{2^{N+1}N(N+1)(N+2)}\,,\\
\delta\gamma_{gq}(N) &= \cf\, \frac{5N^2+7N+4}{2^{N+1}N(N^2-1)}\,,\\
\delta\gamma_{gg}(N)&=\ca\left(
\frac{N+1}{2^{N-1}N(N-1)}+\frac{N+3}{2^{N+1}(N+1)(N+2)}+2B_{1/2}(N+1,0)
\right)\,,
\end{align}
where $B_x(a,b)$ is the incomplete Beta function
\begin{align}
B_x(a,b) = \int_0^x \de t\, t^{a-1}(1-t)^{b-1}\,.
\end{align}
As already mentioned in the main text, we solve the differential equation in Eq.~(\ref{eq: modified dglap}) in two distinct regimes. For \( t < t_d \) (where \( t_d \) is defined in the main text), we evolve all parton flavours, while for \( t > t_d \), the evolution of the heavy-flavour component of \( p_i \) is frozen.
The solution of the differential equation in Mellin space reads
\begin{align}
\label{eq: mellin space p}
    \tilde{p}_{i |i_0}(N,t)=&\; \left[e^{t \Gamma(N)}\right]_{ij } \delta_{ji_0}=  \left[U^{-1}\text{diag}\left(e^{\lambda_1 t},e^{\lambda_2 t},e^{\lambda_3 t}\right)U \right]_{ii_0 },
\end{align}
where \(\lambda_1, \lambda_2, \lambda_3\) are the eigenvalues of \(\Gamma\), \(U\) the matrix that diagonalises \(\Gamma\), and $\tilde{p}_{i_0}(N,0)=\delta_{i i_0}$ is the initial condition in Mellin space.  
The probability distribution in physical space is obtained by performing the inverse Mellin transform of Eq.~\eqref{eq: mellin space p}, integrating in the complex \( N \)-plane along the contour
\begin{align}
\label{eq: Mellin inversion}
    p_{i |i_0}(x,t) = \frac{1}{2\pi i} \int_{c - i\infty}^{c + i\infty} x^{-N} \, \tilde{p}_{i |i_0}(N,t) \, \de N,
\end{align}
where the constant \( c \) is chosen to lie to the right of the rightmost singularity of \( \tilde{p}_{i |i_0}(N,t) \). 
The inversion to physical space has been performed using two distinct numerical methods: the Talbot algorithm~\cite{FT} and the GWR algorithm~\cite{GWR}. These algorithms employ different approaches to evaluate Eq.~(\ref{eq: Mellin inversion}). The Talbot algorithm deforms the integration contour of the inverse Mellin integral to enhance numerical convergence, while the GWR algorithm constructs a discrete sequence of approximations to the inverse transform using the Gaver functional \cite{Gaver_func}, with convergence accelerated by Wynn’s \(\rho\)-algorithm \cite{Wynn_alg}.
The two algorithms yield consistent results for \( x > 1/2 \), whereas the Talbot method becomes unstable in the region \( x < 1/2 \).  
The results obtained from the inverse Mellin transform have been cross-validated against a numerical solution obtained by directly integrating Eq.~(\ref{eq: DGLAP type}).

We conclude this appendix considering the case of the flavour-branch version of the declustering algorithm. The evolution equation in Mellin space becomes in this case particularly simple
\begin{align}
 \frac{\de}{\de t}\tilde{p}_\mathcal{Q}(N,t)= \gamma_{qq}(N)\; \tilde{p}_\mathcal{Q}(N,t),
\end{align}
which leads to
\begin{equation}\label{eq:non-singlet-solution}
   \tilde p_{\mathcal{Q}}(N,t)= e^{\gamma_{qq}(N)t}.
\end{equation}
Because the anomalous dimension appearing in Eq.~(\ref{eq:non-singlet-solution}) is the standard DGLAP one, Mellin inversion can be computed with standard algorithms, such as Talbot.

\section{Soft resummation and dipole evolution} \label{app:dipoles}
In this appendix, we discuss the details of soft resummation. 
Let us start by considering the emission of a soft gluon with momentum $k$ off a massive dipole with momenta $p_i$ and $p_j$. The squared matrix element was given in Eq.~(\ref{eq:w-massive}).
The contribution to the Lund plane density at $\order{\as}$ is~\cite{Lifson:2020gua}
\begin{align}\label{eq:rho-soft-one-emission}
  \rho^{(\text{soft})}_{ij}&=  \as C_{ij}
    \int_0^{Q} k_\perp \de k_\perp \int \de \eta \int_0^{2 \pi}\frac{\de \phi}{4 \pi^2}  w^{ij,k} \Delta \delta(\Delta R-\Delta) k_t \delta( k_t - k_\perp \Delta R)+\order{\as^2},
\end{align}
where $C_{ij}$ is a colour factor.
We have to compute the contribution from all possible dipoles $ij$ that are present in the Born process. For instance, for a $2\to 2$ QCD scattering, we label $1,2$ the incoming legs, $3$ the leg corresponding to the jet we are measuring and $4$ the recoiling one. 
We parametrise the momenta $p_i$ and $k$ as follows:
\begin{subequations}
\begin{align}
    p_1=& \left(E_1,\vec{p}\right), \qquad p_2= \left(E_2,-\vec{p}\right),  \\
    p_3=&  \left(\sqrt{Q^2+m_3^2}\cosh{y_3},Q,0, \sqrt{Q^2+m_3^2}\sinh{y_3} \right), \\
    p_4=&  \left(\sqrt{Q^2+m_4^2}\cosh{y_4},-Q,0, \sqrt{Q^2+m_4^2}\sinh{y_4} \right), \\
    k=& k_\perp \left(\cosh{\eta},\cos{\phi},\sin{\phi},\sinh{\eta}\right),
\end{align}
\end{subequations}
where $E_i= \sqrt{\vec{p}^2+m_i^2}, i=1,2$, $Q$ coincide with the jet transverse momentum, $y_3, y_4$ correspond to the jet and recoiling leg rapidity respectively.
While we wish to keep the masses \( m_3^2 \neq m_4^2 \), we can still choose a frame in which the transverse momenta are balanced. Moreover, even though in collinear factorisation incoming quarks are typically treated as massless, it is possible to outline a calculation with massive initial-state particles using the kinematics described above.
Specifically, we consider the case in which the measured jet is originated by leg 3.
For each dipole configuration, we perform the change of variables:
\begin{align}
    \eta-y_3= \Delta R \cos\psi; \qquad  \phi= \Delta R \sin\psi.
\end{align}
The integrals over $k_\perp$ and $\Delta R$ can be performed trivially using the $\delta$ function. The $\psi$ integration cannot be done in closed form. However, we can consider the quasi-collinear limit, i.e.\ we take $\Delta$ and $\frac{m_i}{Q}$ small, but of the same order. Expanding the result up to order $\order{\Delta^2, \frac{m^2_i}{Q^2}}$ and defining $\Delta_d = \frac{m_3}{Q}$, we find:

\begin{subequations}
\begin{align}
    \rho^{(\text{soft})}_{12}=& \frac{\as C_{12}}{\pi}\left(\Delta^2 +\dots\right)+\order{\as^2},\\
    \rho^{(\text{soft})}_{13}=&\; \frac{\as C_{13}}{\pi} \frac{\Delta^2}{\Delta^2+\Delta^2_d} \left(\frac{\Delta^2}{\Delta^2+\Delta^2_d}+\frac{\Delta^2+\Delta^2_d}{4}+\frac{\Delta^2 \Delta^4_d}{(\Delta^2+\Delta^2_d)^2}+\dots\right)+\order{\as^2}, \\
     \rho^{(\text{soft})}_{23}=& \; \frac{\as C_{23}}{\pi} \frac{\Delta^2}{\Delta^2+\Delta^2_d} \left(\frac{\Delta^2}{\Delta^2+\Delta^2_d}+\frac{\Delta^2+\Delta^2_d}{4}+\frac{\Delta^2 \Delta^4_d}{(\Delta^2+\Delta^2_d)^2}+\dots\right)+\order{\as^2}\\
\rho^{(\text{soft})}_{34}=& \frac{\as \,C_{34}}{\pi} \frac{\Delta^2}{\Delta^2+\Delta^2_d}\left(\frac{\Delta^2}{\Delta^2+\Delta^2_d}+\frac{\Delta^2+\Delta^2_d}{4} \tanh^2{\frac{Y}{2}}
+\frac{\Delta^4_d\Delta^2}{(\Delta^2+\Delta^2_d)^2}+\dots\right)+\order{\as^2}, \\
 \rho^{(\text{soft})}_{14}=&\;  \frac{\as C_{14}}{\pi} \left(\Delta^2 \frac{e^{Y}}{4 \cosh^2 \frac{Y}{2}}+\dots\right)+\order{\as^2}, \\
\rho^{(\text{soft},1)}_{24}=& \;  \frac{\as C_{24}}{\pi} \left(\Delta^2 \frac{e^{Y}}{4 \cosh^2 \frac{Y}{2}}+\dots\right)+\order{\as^2},
\end{align}
where the dots indicate higher order contribution $\mathcal{O}\left(\Delta^4,\frac{m_i^4}{Q^4},\Delta^2 \frac{m_i^2}{Q^2}\right), i=1,2,3,4$ and $Y=y_3-y_4$.
As expected, mass effects are genuine power corrections in the dipoles that do not involve the jet. 
We observe that both  the first two contributions coincide with the soft limit of the quasi-collinear splitting function in Eq.~(\ref{eq: rho QC soft}).
The massless limit is smoothly recovered as $\Delta_d\to 0$.
\end{subequations}

We now move our discussion to parametrisation of the emissions' kinematics, which is a crucial ingredient of the dipole shower that we use to resum soft logarithms. 
Let us first introduce a few kinematic quantities:
\begin{subequations}\begin{align}
  s_{ij} & = (p_i+p_j)^2, & \lambda & = (1-\mu_i-\mu_j)^2-4\mu_i\mu_j, \\
  p_i^2 &= m_i^2,  & p_j^2 & = m_j^2, \\
  \mu_i & = \frac{m_i^2}{s_{ij}}, & \mu_j & = \frac{m_j^2}{s_{ij}} .
\end{align}
\end{subequations}
In the presence of massive partons, we have some freedom regarding the choice of reference momenta to perform the Sudakov parametrisation of the emission momentum $k$. For instance, the Sudakov decomposition can be taken either in terms of the
(potentially) massive vectors directly, or in terms of lightlike
vectors $(n_i,n_j)$ defined to be aligned with $p_i$ and $p_j$ in the
dipole rest frame, with normalisation to be specified later:
\begin{equation}
  k^\mu = \zeta_i p_i^\mu + \zeta_j p_j^\mu + k_\perp^\mu = z_i n_i^\mu + z_j n_j^\mu + k_\perp^\mu.
\end{equation}
The four vector $k_\perp^\mu$ is a spacelike vector that is orthogonal to $p_i, p_j, n_i, n_j$ and $k_\perp^2= -\kt^2$.
While we have tried both approaches, we find that the second choice is preferred because one can achieve full phase-space coverage with $0\le z_{i,j}\le 1$, while with the first option, full phase-space coverage can only be achieved by allowing $\zeta_{i,j}$ to be negative.  
Thus, we use the lightlike vectors $n_i, n_j$ and the momentum fractions $z_{i,j}$. In general, the lightlike vectors can be written as
\begin{subequations}\begin{align}
  n_i & = y_i[p_i-x_i (p_i+p_j)],\\
  n_j & = y_j[p_j-x_j (p_i+p_j)].
\end{align}\end{subequations}
The coefficients $x_{i,j}$ are fixed by requiring that $n_{i,j}^2=0$, giving
\begin{subequations}\begin{align}
  x_i & = \frac{1}{2}(1+\mu_i-\mu_j-\sqrt{\lambda}),\\
  x_j & = \frac{1}{2}(1-\mu_i+\mu_j-\sqrt{\lambda}).
\end{align}\end{subequations}
The normalisation factors $y_{i,j}$ can be adjusted so that the Sudakov variables $z_i$ and $z_j$ coincide with  the momentum fractions of the emission with momentum $k$ with respect to $p_i$ and $p_j$. Indeed we have 
\begin{equation}
  \frac{n_j\cdot k}{n_j\cdot p_i}= \frac{z_i (1-x_j)}{y_i (1-x_i-x_j)},
  \qquad\text{and}\qquad
  \frac{n_i\cdot k}{n_i\cdot p_j}= \frac{z_j (1-x_i)}{y_j (1-x_i-x_j)}.
\end{equation}
Thus, by choosing 
\begin{equation}
  y_i = \frac{1-x_j}{1-x_i-x_j},
  \qquad\text{and}\qquad
  y_j = \frac{1-x_i}{1-x_i-x_j},
\end{equation}
we obtain
\begin{equation}
  z_i = \frac{n_j\cdot k}{n_j\cdot p_i},
  \qquad\text{and}\qquad
  z_j = \frac{n_i\cdot k}{n_i\cdot p_j}.
\end{equation}
In practice, we therefore parametrise emissions using
\begin{equation}
  z_i = \frac{k_t}{Q_{ij}} e^{\eta},
  \qquad\text{and}\qquad
  z_j = \frac{k_t}{Q_{ij}} e^{-\eta}.
\end{equation}
Note that this is the same functional form as for the massless case~\cite{Lifson:2020gua}, but with $Q_{ij}^2 = 2n_i.n_j = (1-x_i)(1-x_j)s_{ij}$ and the phase-space
boundaries
\begin{equation}
  k_t\le Q_{ij},
  \qquad
  |\eta| \le \log\frac{Q_{ij}}{k_t}.
\end{equation}
With these variables, the eikonal factor defined in Eq.~\eqref{eq:w-massive} can be shown to
take the form
\begin{align}
w^{ij,k} = \frac{2}{k_t^2}\left[\frac{1-x_i-x_j}
    {(1-x_i+x_ie^{2\eta})(1-x_j+x_je^{-2\eta})}\right]^2.
\end{align}

\section{Details of the fixed-order calculation} \label{app:e+e-lo}
In this section we derive the order $\order{\as}$ calculation for the Lund plane density in $e^+ e^-$ collisions. We consider:\footnote{For simplicity we consider only the $\gamma^*$ induced process. Radiative corrections to the Born process $e^+ e^- \to Z \to b \bar b$ differ by power correction in $m^2/s$.}
\begin{equation}
\label{eq: qed process-app}
e^-(p_a)\, e^+(p_b) \to\gamma^*\to  b(p_1) \, \bar b(p_2)\,  g(p_3),
\end{equation}
with $p_1^2=p_2^2=m^2$, $p_a^2=p_b^2=p_3^2=0$, $s=2 p_a \cdot p_b$. The kinematics of this process is usually defined in terms of dimensionless ratios:
\begin{align}
    x_i=\frac{2 p_i \cdot (p_a+p_b)}{s},\quad  i=1,2,3.
\end{align}
 Where in this appendix, as customary, the variables $x_i$ represent
 the energy fractions with respect to the energy of the incoming beam,
 in the centre-of-mass frame, $x_i=\frac{2 E_i}{\sqrt{s}}$, such that $x_1+x_2+x_3=2$. As already stated, $Q= \frac{\sqrt{s}}{2}$ represents the jet hard scale.  
 The double differential distribution reads:
\begin{align}
  \frac{1}{\sigma_0}\frac{\de^2\sigma}{\de x_1 \de x_2}=&\frac{\cf\as}{2\pi}\frac{1}{\beta}\Bigg[
   \frac{2(x_1+x_2-1)-4\mu}{(1-x_1) (1-x_2)}-2\mu\left(\frac{1}{(1-x_1)^2}+\frac{1}{(1-x_2)^2}\right)\nonumber\\
  +&\frac{1}{1+2\mu} \frac{(1-x_1)^2+(1-x_2)^2}{(1-x_1)(1-x_2)}\Bigg],  \quad \mu= \frac{ m^2}{s}, \quad \beta = \sqrt{1-4\mu}
\end{align}
where $\sigma_0$ is the Born cross section.
 We divide each event into two hemispheres and we measure the Lund plane density of each of them. With just one emission, this amounts to compute
\begin{align} \label{eq:density-lo-start}
    \widetilde{\rho}_\mathcal{Q}^\text{(f.o.)}(\eta,\kt)=& \frac{\kt}{2 \sigma_0 \cosh^2 \eta} \,  \int_{2\sqrt{\mu}}^1 \de x_1 \int_{x_2^-}^{x_2^+}\de x_2 \,\frac{\de^2 \sigma}{\de x_1 \de x_2} \nonumber\\ & \Bigg[
    \Theta(\theta_{13}< \min (\theta_{23},\theta_{12}))
    \delta(\tanh\eta-\cos\theta_{13}) \delta \left(k_t -\frac{\sqrt{s}}{2}\frac{ \min(x_1,x_3)}{\cosh \eta}\right)\nonumber\\
    &+ \Theta(\theta_{23}< \min (\theta_{13},\theta_{12}))
    \delta(\tanh \eta-\cos\theta_{23}) \delta \left(k_t -  \frac{\sqrt{s}}{2}\frac{\min(x_2,x_3)}{\cosh \eta}\right)\nonumber\\
   &+ \Theta(\theta_{12}< \min (\theta_{13},\theta_{23}))
    \delta(\tanh \eta-\cos\theta_{12}) \delta \left(k_t - \frac{\sqrt{s}}{2}\frac{\min(x_1,x_2)}{\cosh\eta}\right)
    \Bigg],
\end{align}
where the factor of $1/2$ accounts for the fact that we have two jets. 
In Eq.~(\ref{eq:density-lo-start}), we have introduced~\cite{Oleari:1997az}
\begin{equation}
    x_2^{\pm}=\frac{  (2-x_1)(2\mu + (1-x_1)\pm  (1-x_1) \sqrt{x_1^2-4\mu }}{2 (1-x_1+\mu)}.
\end{equation}
The $\Theta$-functions defines 3 angular regions, depending on which pair of partons is the closest in angle. 
The angles $\theta_{ij}$ are expressed in terms of the energy fraction
$x_i$ in the centre of mass frame:
\begin{align}
    \cos \theta_{13}&= \frac{1}{v_1}\left(1-\frac{2(1-x_2)}{x_1 x_3}\right), &
    \cos \theta_{23}&= \frac{1}{v_2}\left(1-\frac{2(1-x_1)}{x_2 x_3}\right), \nonumber \\
   \cos \theta_{12} &= \frac{1}{v_1 v_2}\left(1-2\frac{1-x_3-8\mu}{x_1 x_2}\right), &
   v_{1,2}&= \sqrt{1- \frac{4\mu}{x^2_{1,2}}}.
\end{align}
The $k_t$ $\delta$-function further subdivides each region in two. However, of the 6 regions, only 3 are independent. The case of $bg$ being the closest pair is equivalent to the $\bar b g$. Furthermore, the region in which $b\bar b$ is the closest pair, is symmetric upon the exchange $x_1 \leftrightarrow x_2$.
We note that, in each region, we have two integrals and two $\delta$-functions, so the only nontrivial exercise is to work out the Jacobian factors. Finally, the calculation of the Lund plane plane density for massless quarks can be simply obtained by setting $\mu=0$ in Eq.~\eqref{eq:density-lo-start}.
\addcontentsline{toc}{section}{References}

\bibliographystyle{jhep}
\bibliography{biblio}
\end{document}